\documentclass[aps,twocolumn,showpacs,amsmath,amssymb,floatfix,superscriptaddress]{revtex4-1}

\usepackage{color}
\usepackage{bbm}
\usepackage{graphicx}% Include figure files
\usepackage{dcolumn}% Align table columns on decimal point
\usepackage{bm}% bold math
\usepackage{array}
\usepackage{float}
\usepackage{supertabular}
\usepackage{longtable}
\usepackage{mathrsfs}
\usepackage{txfonts}
\usepackage{wasysym}
\usepackage[T1]{fontenc}
\usepackage[latin1]{inputenc}
\usepackage{amssymb}
\usepackage[usenames,dvipsnames]{xcolor}
\usepackage{amsmath}
\usepackage{cases}
\usepackage{amstext}
\usepackage{latexsym}
\usepackage[colorlinks=true,citecolor=Cerulean,linkcolor=RubineRed,urlcolor=Cerulean]{hyperref}
\usepackage{enumitem}

\begin{document}
\title{
Nonlinear quantum metrology of many-body open systems 
}

\author{M. Beau}
\affiliation{Department of Physics, University of Massachusetts, Boston, MA 02125, USA}
\author{A. del Campo}
\affiliation{Department of Physics, University of Massachusetts, Boston, MA 02125, USA}

\newcommand{\be}{\begin{equation}}
\newcommand{\ee}{\end{equation}}
\newcommand{\bea}{\begin{eqnarray}}
\newcommand{\eea}{\end{eqnarray}}

\newcommand{\bmat}{\begin{pmatrix}}
\newcommand{\emat}{\end{pmatrix}}

\def\S{\mathcal{S}}
\def\E{\mathcal{E}}
\def\q{{\bf q}}

\def\G{\Gamma}
\def\L{\Lambda}
\def\la{\lambda}
\def\g{\gamma}
\def\al{\alpha}
\def\s{\sigma}
\def\e{\epsilon}
\def\k{\kappa}
\def\ve{\varepsilon}
\def\l{\left}
\def\r{\right}
\def\te{\mbox{e}}
\def\d{{\rm d}}
\def\t{{\rm t}}
\def\K{{\rm K}}
\def\N{{\rm N}}
\def\H{{\rm H}}
\def\la{\langle}
\def\ra{\rangle}
\def\om{\omega}
\def\Om{\Omega}
\def\vep{\varepsilon}
\def\wh{\widehat}
\def\tr{{\rm Tr}}
\def\da{\dagger}
\def\iz{\left}
\def\zi{\right}
\newcommand{\beq}{\begin{equation}}
\newcommand{\eeq}{\end{equation}}
\newcommand{\beqa}{\begin{eqnarray}}
\newcommand{\eeqa}{\end{eqnarray}}
\newcommand{\intf}{\int_{-\infty}^\infty}
\newcommand{\into}{\int_0^\infty}

\begin{abstract}
We introduce  general bounds for the parameter estimation error in nonlinear quantum metrology of many-body open  systems in the Markovian limit.
Given a $k$-body Hamiltonian and $p$-body Lindblad operators, the estimation error of a Hamitonian parameter using a Greenberger-Horne-Zeilinger (GHZ) state as a probe is shown to scale as $N^{-\left(k-\frac{p}{2}\right)}$, surpassing the shot-noise limit for $2k>p+1$.
Metrology equivalence between initial product states and maximally entangled states is established for $p\geq 1$. 
We further show that one can estimate the system-environment coupling parameter with precision $N^{-\frac{p}{2}}$, while  many-body decoherence enhances the precision to $N^{-k}$ in the noise-amplitude estimation of a fluctuating $k$-body Hamiltonian.  
For the long-range Ising model we show that the precision of this parameter beats the shot-noise limit when the range of interactions is below a threshold value.
\end{abstract}

\maketitle

\begin{figure}
\includegraphics[width= 1\columnwidth]{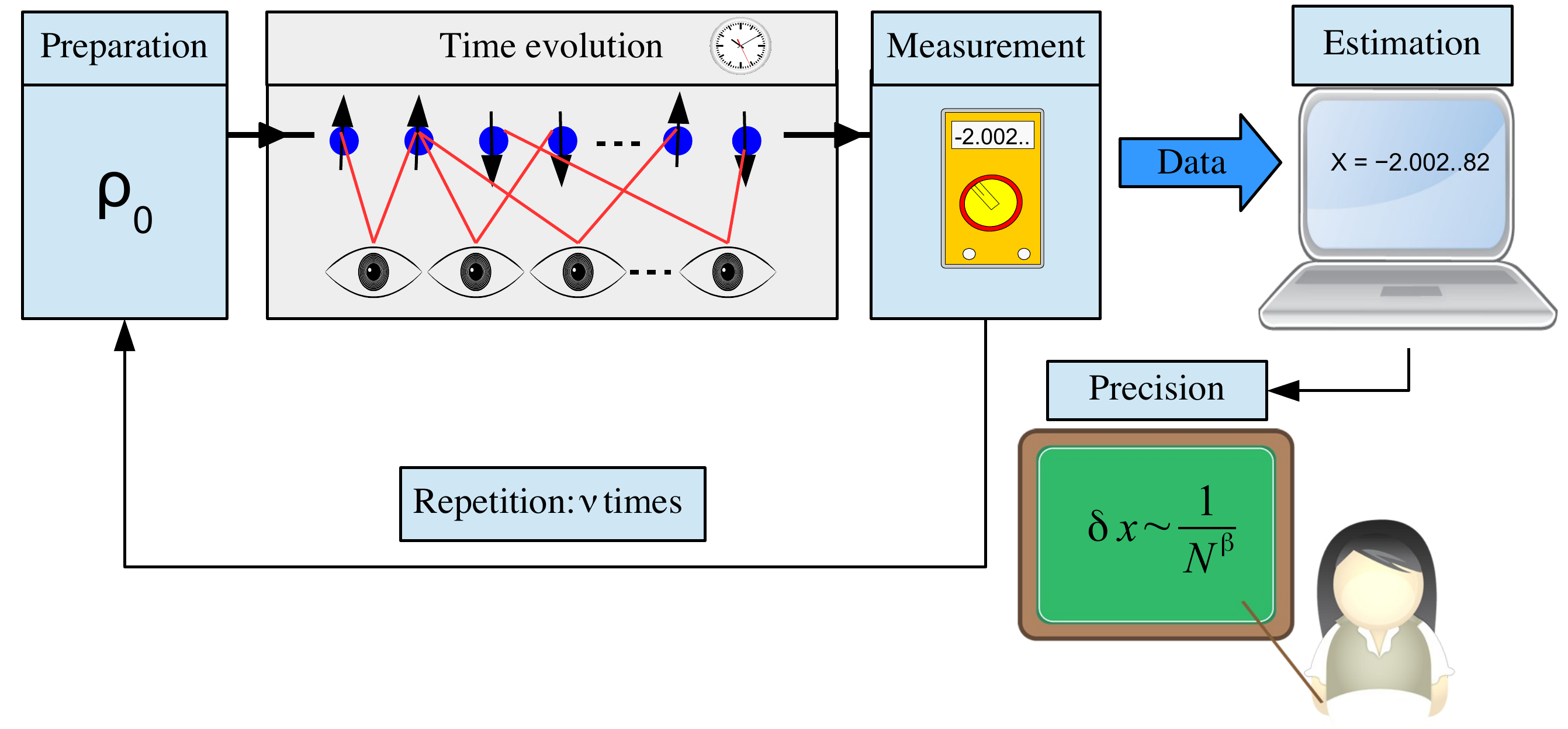}\\
\caption{{\bf Quantum metrology protocol of a many-body open quantum system. } 
A protocol with $N$ two-level systems evolving under a long-range dissipator with 2-body Lindblad operators accounting for the role of a monitoring environment (represented by the eyes) is illustrated. After repeating the process $\nu$ times, the value of a parameter $x$ is estimated with precision $\delta x$ that scales as $N^{-\beta}$. 
%The purpose of this letter is to compute the exponent $\beta$ for different initial states $\rho_0$. 
 \label{Fig1}}
\end{figure}

Quantum metrology exploits quantum resources to enhance the precision in the parameter estimation of a physical system \cite{Giovannetti04,Giovannetti11}. It has broad applications across different fields ranging from gravitational-wave detectors \cite{Caves81,LIGO11} to atomic spectroscopy and time-frequency standards \cite{Wineland92,Bollinger96,Giovannetti01}. The general protocol consists of preparing $N$ particles in an initial state (e.g., with entanglement or squeezing) that evolves in time before detection, see Fig. \ref{Fig1}. This process is repeated many times to collect statistics and provide an estimation of the uncertain parameter.    
The estimation error depends on the resources for the probe and the protocol, and specifically,  on the number of particles $N$, the duration $t$ of each repetition, and the number of repetition $\nu=T/t$ where $T$ is the total time of the protocol. The mathematical framework used to evaluate the (unbiased) estimation error is based on the quantum Cram\'{e}r-Rao bound (QCRB) that depends on the quantum Fisher information (QFI) \cite{Helstrom76,Holevo82,Wiseman10,Braunstein94,Braunstein96}. With classical resources (uncorrelated product states), the central limit theorem shows that the estimation error of a parameter $x$ scales as the so-called shot-noise limit (or standard limit) $\delta x \sim 1/\sqrt{N\nu}$ for large $N$ \cite{Braunstein94,Braunstein96}. Surpassing this limit is the challenge of quantum metrology. Using as a probe  $N$ noninteracting particles prepared in a Greenberger-Horne-Zeilinger (GHZ) state that maximizes the variance of the Hamiltonian, it is well established that the error beats the standard limit and scales as $1/(N\sqrt{\nu})$, referred to as the Heisenberg limit. Unfortunately, the presence of noise in the system destroys any gain of precision \cite{Huelga97}. Recent progress has been made in noisy quantum metrology to achieve optimal sensitivity \cite{Giovannetti11,Demkowicz14,Demkowicz12,Escher11,Escher11bis,Escher12,Smirne16,Chin12,Chaves13,Dur14}.
Nonlinear estimation strategies open yet new frontiers \cite{Boixo07,Giovannetti11} and progress has been made towards a general theory \cite{Boixo08,Hall12,Alipour14,Alipour15,Yousefjani16} and its applications, e.g., in Bose-Einstein condensates \cite{Rey07,Choi08,Boixo08b,Boixo09} and optical atomic clocks \cite{MartinThesis,Martin13,Ludlow15}. 

In this Letter, we focus on the estimation of Hamiltonian and bath-coupling parameters in nonlinear quantum metrology of many-body open systems. Specifically, we establish new bounds for the QFI of Markovian open quantum systems and propose a general method to derive sharp QCRB for pure dephasing dissipators. For initial states maximizing the variance of the Hamiltonian, we show that the \textit{Zeno time} -set by the inverse of energy fluctuations that govern the short-time quantum decay of the probe state under unitary dynamics- characterizes the scaling of the error on the Hamiltonian coupling parameter $x_1$. On the other hand, for open quantum systems the short-time decay is linear in time due to decoherence. The so-called \textit{decoherence time} \cite{CBCdC16} that sets the decay rate  also defines the scale of the optimal interrogation time. Therefore we show that the precision of the estimation of $x_1$ is reduced as the decoherence time decreases while it is enhanced for the environment-system coupling parameter $x_2$.  
%, as well as the estimation error for the environment-system coupling parameter $x_2$.
For a $k$-body Hamiltonian and $p$-body Lindblad operators, the error is shown to scale as $\delta x_1\sim N^{-\left(k-\frac{p}{2}\right)}$ and $\delta x_2\sim N^{-\frac{p}{2}}$ respectively, which may surpass the shot-noise limit. It follows that the the super-Heisenberg limit \cite{Boixo08b} is robust for $k-\frac{p}{2}>1$.  In the most general situation, we prove the metrology equivalence between initial product states and the  GHZ states by extending the results to the case of a many-body dissipator. We apply our method to the long-range Ising chain, and propose a scenario to estimate the amplitude of a stochastic noise in a fluctuating many-body Hamiltonian system.  We hope our unifying framework of non-linear quantum metrology will open new frontiers of parameter estimation for trapped ions \cite{Richerme14,Jurcevic14} and cold atoms \cite{Bloch08} with application to the design of optical lattice clocks with many-body spin systems \cite{MartinThesis,Martin13,Ludlow15}.
  
\textit{Quantum Cram\'er-Rao bound for open quantum systems.} % General method
The evolution of an initial state $\rho_{0}$ under Markovian dynamics satisfies the following general master equation  \cite{Lindblad76,Breuer02,Lidar06}
\begin{align}\label{MasterEq1}
\frac{d}{dt}\rho(t) &= \mathcal{L}(\rho(t)) = -\frac{ix_1}{\hbar}\left[H,\rho(t)\right]
\\ \nonumber &+\sum_{\alpha=2}^{d+1}x_\alpha\left(L_\alpha\rho(t)L_\alpha^\dagger-\frac{1}{2}L_\alpha^\dagger L_{\alpha}\rho(t)-\frac{1}{2}\rho(t)L_\alpha^\dagger L_\alpha\right),\ 
\end{align}
where the coupling constant $x_\alpha \geq 0$ for $\alpha\geq 2$, and where $d\leq 4^N-1$ when $\rho_0$ describes $N$ two-level systems.

We shall consider the estimation of the coupling constants $x_\alpha,\ \alpha\geq 1$ and elucidate the effect of the dissipator on the estimate of the parameter $x_1$ representing the strength of the Hamiltonian. 
The theoretical framework of quantum metrology \cite{Helstrom76,Holevo82,Wiseman10,Braunstein94,Braunstein96} bounds the estimation error using the QCRB 
\begin{equation}\label{QCRB}
\delta x_\alpha \geq 1/\sqrt{\nu \mathfrak{F}_{x_\alpha}(t)}\ ,
\end{equation}  
where $\delta x_\alpha = \sqrt{\la ((x_\alpha)_{\text{est}}-x_\alpha)^2\ra}$ assuming an unbiased estimate, $\nu$ is the number of repetitions of a measurement after a time $t$ and $\mathfrak{F}_{x_\alpha}(t)$ is the quantum Fisher information (QFI) given by
\begin{equation}
\mathfrak{F}_{x_\alpha}(t)\equiv \text{Tr}\left[\rho(t)\mathfrak{L}_{x_\alpha}^2(t)\right]\ ,
\end{equation}
where $\mathfrak{L}_{\gamma_\alpha}^2(t)$ corresponds to the symmetric logarithmic derivative formally defined as $\partial_{x_\alpha}\rho(t)=\left(\mathfrak{L}_{x_\alpha}(t)\rho(t)+\rho(t)\mathfrak{L}_{x_\alpha}(t)\right)/2$. For closed systems,  the simplified form of the $\mathfrak{L}_{\gamma_\alpha}^2(t)$ yields the expression for QFI $4t^2 \Delta_\rho H^2/\hbar$ for a pure state $\rho(0)$, where $H$ is the generator of the dynamics, $\Delta_\rho H^2\equiv \la H^2\ra_\rho -\la H\ra_\rho^2$, and $\la \bullet\ra_\rho\equiv \text{Tr}(\bullet\rho)$ denotes the quantum average with respect to the density matrix $\rho$.  
Under Markovian dynamics, we use the spectral decomposition of the density matrix $\rho(t)=\sum_{n}\xi_n|\xi_k\ra\la \xi_n|$, in terms of its orthonormal eigenvectors $|\xi_n\ra$ and associated eigenvalues $\xi_n$. The explicit expression of the QFI reads
\begin{equation}\label{QFI1}
\mathfrak{F}_{x_\alpha}(t)=\sum_{n,n': \xi_n+\xi_{n'}\neq 0}4\xi_n \frac{|\la \xi_n| \partial_{x_\alpha}\rho(t)|\xi_{n'}\ra|^2}{(\xi_n+\xi_{n'})^2} \ .
\end{equation}
An upper bound for the QFI associated to $x_1$ in a closed system was derived in \cite{Braunstein94}, see also \cite{Braunstein96} and references therein. As we show in \cite{SM},
 its extension to open quantum system is given by\begin{equation}\label{QFIbound1}
c_m \frac{4t^2\Delta_{\rho} H^2}{\hbar^2} \leq \mathfrak{F}_{x_1}(t)\leq c_M \frac{4t^2\Delta_{\rho} H^2}{\hbar^2} \ ,
\end{equation}
where $\Delta_{\rho} H^2 = \la H^2 \ra_{\rho} - \la H \ra_{\rho}^2$ is the variance of the Hamiltonian in the state $\rho(t)$, and the two coefficients take the form 
\begin{equation}\label{Coef}
c_{m}\equiv \left(\frac{1-r}{1+r}\right)^2\ ,\ c_{M}\equiv \left(\frac{\xi_{M}-\xi_{m}}{\xi_{M}+\xi_{m}}\right)^2\ ,
\end{equation}
with $\xi_{M(m)}$ being the maximum (minimum) eigenvalue of the density matrix $\rho$, and $r=\max\limits_{n<n'}\left(\frac{\xi_n}{\xi_{n'}}\right)$. For unitary dynamics, these two coefficients are time-independent and are equal to one for a pure state (as $\xi_M=1$ and $\xi_{n}=0,\ n\neq M$), leading to the QCRB $\delta x_1 \geq \hbar/(\Delta_{\rho_0}H\sqrt{\nu})$. 
% We note that the bound \eqref{QFIbound1} is valid for mixed states.
For Markovian open quantum system, see equation \eqref{MasterEq1}, the  eigenvalues $\xi_n$ of the density matrix are modified by the presence of the dissipator, and the coefficients $c_{M},\ c_m$ are time-dependent. 
From the bounds \eqref{QFIbound1}, we deduce that the decay of QFI has two possible origins: dissipation and dephasing. The first is associated with the loss of energy in the system
$
\frac{d}{dt}\la H \ra_{\rho} = -\sum_{\alpha}x_{\alpha}\text{tr}\left(\rho\left[L_\alpha,\left[L_\alpha,H\right]\right]\right),
$   
while the second is encoded in the two coefficients $c_{m},\ c_{M}$, see equation \eqref{Coef}. 
We focus on the role of dephasing by taking $[H,L_\alpha]=0$ so that the mean value and the variance of the energy are conserved in time. 

\textit{Many-body Hamiltonian and Lindblad operators.} % QCRB symmetrized L operator: main results 
We consider the general master equation \eqref{MasterEq1} with a symmetrized $k$-body Hamiltonian and a $p$-body Hermitian Lindblad operators
\begin{equation}\label{kbodyHpbodyLGeneral}
\H=\sum_{i_1<\cdots <i_k}H_{i_1,\cdots, i_k}\ ,\ \text{and}\  L_\alpha = L_{i_1,\cdots,i_p}\ ,
\end{equation}
where the sum runs over all k-tuples $\left(i_1,\cdots ,i_k\right)$ with $i_1<\cdots <i_k$. The coupling constant $x_\alpha$ are considered to be equal with each other $x_{\alpha}=x_2$. As a shorthand we denote the $k$-body Hamiltonian operators by $H_{\mu}$ and the $p$-body Lindblad operator by $L_{\nu}$, where $\mu$ and $\nu$ denote vectors in the set if k-tuples $\mathfrak{S}_k\equiv\left\{(i_1,\cdots,i_k)\right\}$ and $\mathfrak{S}_p\equiv\left\{(i_1,\cdots,i_p)\right\}$ respectively, and where the sum over the $n$-tuples reads $\sum_{\nu\in\mathfrak{S}_n} = \sum_{i_1<\cdots <i_n}$ . We consider that all the $k$- and $p$-body operators commute with each other $[H_{\mu},H_{\mu'}]=0,\ [L_\nu,L_{\nu'}]=0,\ [H_{\mu},L_{\nu'}]=0$. 
The master equation \eqref{MasterEq1} takes the form
\begin{multline}\label{MasterEq2}
\frac{d}{dt}\rho(t)=\mathcal{L}(\rho(t))= \\ -\frac{ix_1}{\hbar} \sum_{\mu\in\mathfrak{S}_k} [H_\mu,\rho(t)] 
+ x_2\sum_{\nu\in\mathfrak{S}_p}\left( L_{\nu}\rho L_{\nu}-\frac{1}{2}L_{\nu}^2\rho-\frac{1}{2}\rho L_{\nu}^2\right)\ .
\end{multline}
First, we specify the density matrix elements $\rho_{ij}(t)=\la e_i|\rho(t)|e_j \ra$, where $\left\{|e_{j}\ra\right\}$ is the orthonormal basis of $N$-particle system, see \cite{SM}. We choose the eigenbasis of the $k$-body Hamiltonian and $p$-body Lindblad operators so that $H_{\mu}=\sum_{i}\epsilon_{i}^{(\mu)}|e_i \ra\la e_i|$ and $L_\nu=\sum_{i}\lambda_{i}^{(\nu)}|e_i \ra\la e_i|$. The master equation \eqref{MasterEq2} leads to  
$
\dot{\rho}_{ij}(t)=\left(-\frac{ix_1}{\hbar}\epsilon_{ij}-\frac{x_2}{2}\lambda_{ij}^2\right)\rho_{ij}(t)\ ,
$  
with
$
\epsilon_{ij}\equiv\sum_{\mu}\left(\epsilon_i^{(\mu)}-\epsilon_j^{(\mu)}\right)\ ,\ \lambda_{ij}^2\equiv\sum_{\nu}\left(\lambda_i^{(\nu)}-\lambda_j^{(\nu)}\right)^2\ .
$
Thus, the time-evolution of the density matrix reads
\begin{equation}\label{rhot}
\rho_{ij}(t) = \rho_{ij}(0)e^{\left(-\frac{ix_1}{\hbar}\epsilon_{ij}-\frac{x_2}{2}\lambda_{ij}^2\right) t}\ .
\end{equation}
%The results obtained so far are valid for an arbitrary initial state. Equations \eqref{QFIbound1} and \eqref{rhot} provide a method to solve analytically and numerically the metrology problem for systems governed by the master equation \eqref{MasterEq2}.
In what follows, we solve the parameter-estimation problem for initial maximally-entangled and product states, and show the applications to the long-range Ising model and to a Hamiltonian with fluctuating many-body interactions, see Fig. \ref{Fig1} for an illustration of the metrology protocol. 

\textit{States maximizing the many-body energy variance.} Consider as a probe the state that maximizes the variance of the Hamiltonian $H=\sum_{\mu} H_{\mu}$, that is $\rho_{0}=|\Psi_0\ra\la\Psi_0|$, with $|\Psi_0\ra = \left(| E_m\ra+|  E_M\ra\right)/\sqrt{2}$ where $ |E_{m(M)}\ra$ is the eigenstate associated with the minimum (maximum) eigenvalue $E_{m(M)}$ of the Hamiltonian $H$. We denote $\lambda_{m(M)}^{(\nu)}$ the eigenvalue of the $p$-body operators $L_{\nu}$ associated with the state $|E_{m(M)}\ra$. Using the exact evolution of the density matrix  \eqref{rhot},  we find in \cite{SM} that the  fidelity is given by
%\begin{subequations}
%\begin{equation}\label{purityGHZ}
%p(t) = \frac{1}{2}\left(1+e^{-4 t/\tau_D}\right) \ ,
%\end{equation}
\begin{equation}\label{FidelityGHZ}
F(t) = \text{tr}\left(\rho_t\rho_0\right) = \frac{1}{2}\left(1+e^{-2t/\tau_D}\cos{\left(t/\tau_Z\right)}\right) \ ,
\end{equation}
where the Zeno time $\tau_Z$ and the decoherence time $\tau_D$ are given by
\begin{equation}\label{tau}
\tau_Z \equiv \frac{\hbar}{2x_1\Delta H} \ \  ,\ 
\tau_D \equiv \frac{1}{x_2\sum_{\nu}\Delta L_\nu^2} \ . 
\end{equation}
Here, the variance of the Hamiltonian is related to its seminorm $\|H\|$, see \cite{Boixo07},
$
\Delta H = \frac{1}{2}\| H\|=\frac{1}{2}\left(E_M-E_m\right)
$
and  the decoherence time is set by  the square of the variance of the Lindblad operator \cite{CBCdC16} that can be computed explicitly  
$
\Delta L_\nu^2=\frac{1}{4}\left(\lambda_M^{(\nu)}-\lambda_m^{(\nu)}\right)^2.
$
It is well known that these two time-scales play a role in the dynamics of the system. The Zeno time provides the quantum speed limit to the fidelity decay of the closed system (see master equation \eqref{MasterEq2} with $x_2=0$) while the decoherence time sets the dephasing-induced short-time decay rate of the fidelity $F(t)\approx 1-\frac{t}{\tau_D}$, see equation \eqref{FidelityGHZ}. As we show in \cite{SM}, these time scales also govern the scaling of the estimation errors. 
Indeed, the QFI \eqref{QFI1} depends explicitly on $\tau_Z$ and $\tau_D$ 
\begin{equation}\label{QFIx}
\mathfrak{F}_{x_1}(t) = \frac{t^2}{x_1^2\tau_Z^2} e^{-4 t/\tau_D}\ \ ,\ \  
\mathfrak{F}_{x_2}(t) =\frac{t^2}{x_2^2\tau_D^2} \frac{e^{-4 t/\tau_D}}{1-e^{-4 t/\tau_D}}\ .
\end{equation}
Notice that $\mathfrak{F}_{x_1}(t)$ in equation \eqref{QFIx} can also be derived using the two bounds \eqref{QFIbound1} as the coefficients $c_m$ and $c_M$ are both equal to $e^{-4 t/\tau_D}$ and $\Delta_{\rho_0}H = \hbar/(2\tau_Z)$ by definition \eqref{tau}.  
Equation \eqref{QCRB} gives the QCRB that read $\delta x_1\geq (x_1\tau_Z/(t\sqrt{\nu}))e^{2 t/\tau_D}$ and $\delta x_2\geq (x_2\tau_D/(t\sqrt{\nu}))\sqrt{e^{4 t/\tau_D}-1}$, where $\nu$ is the number of repetitions. It follows that $\tau_Z$ and $\tau_D$ directly determine the amplitude of the estimation error of $x_1$ and $x_2$.  

A metrology protocol identifies a suitable short interrogation time for each measurement and data is collected for a total time $T$. To estimate parameter $x_i$, the optimal interrogation time maximizes the associated QFI $\mathfrak{F}_{x_i}$ with respect to $x_i$. Under (\ref{MasterEq2}), it is set by the decoherence time up to a constant $t_{i}=\mu_i \tau_D$ where $\mu_1=0.5$ and $\mu_2\approx 0.40(1)$ for the estimation of $x_1$ and $x_2$, respectively (see \cite{SM}). As a result, we find that the sensitivity scales as   
\begin{equation}\label{EstError}
\ \ \ \ \delta x_1 \geq \frac{x_1}{\kappa_1}\frac{\tau_z}{\sqrt{ T\tau_D}}\ \ ,\ 
\delta x_2 \geq \frac{x_2}{\kappa_2}\sqrt{\frac{\tau_D}{ T}} \ ,
\end{equation}
with $\kappa_1 = \mu_1^{1/2}e^{-2\mu_1}$ and $\kappa_2 = \mu_2^{1/2}\left(e^{4\mu_2}-1\right)^{-1/2}$.
For small time $t\ll \tau_D$, the QFI $\mathfrak{F}_{x_2}(t)$ in equation \eqref{QFIx} grows linearly in time $\mathfrak{F}_{x_2}(t) = t/(4x_2^2\tau_D)$. This implies that the QCRB \eqref{QCRB} scales as $x_2\sqrt{\tau_D / (t \nu)}$ which has the same scaling as $\delta x_2$ in equation \eqref{EstError} if the repetition number $\nu$ is not too large. By contrast, for the same time scale $t\ll \tau_D$ the QFI for $x_1$ in equation \eqref{QFIx} reads $\mathfrak{F}_{x_1}(t) = t^2/(x_1^2\tau_Z^2)$, leading to the same QCRB that is obtained without decoherence, i.e., $x_1\tau_z/(t\sqrt{\nu})$.
For the dephasing model \eqref{MasterEq2}, we can easily show that the joint estimation strategy \cite{Helstrom76,Holevo82,Humphreys13,Cheng14} does not improve the precision of the measurement of the two parameters $x_1$ and $x_2$.  Joint estimation protocols offer however exciting prospects when this is no longer the case, e.g., in the presence of dissipation.

For instance, consider as an initial state the GHZ-state of a $k$-body Hamiltonian
\begin{equation}\label{Hproduct}
H=\sum_{i_1<\cdots<i_k}H_{i_1}\otimes\cdots\otimes H_{i_k}\ .
\end{equation} 
The GHZ-state reads 
$ 
|\Psi_0\ra = \frac{1}{\sqrt{2}}\left(|\varepsilon_m \ra^{\otimes N} + |\varepsilon_M\ra^{\otimes N}\right) 
$
where $ |\varepsilon_{m(M)} \ra $ is the eigenstate associated with the minimum (maximum) eigenvalue $\varepsilon_{m(M)}$ of the single-body Hamiltonian $H_i$. The $k$-body minimum (maximum) eigenvalue is $E_{m(M)}={N\choose k}\times(\varepsilon_{m(M)})^k$, and we assume that it is not degenerate, more specifically that $E_{m}\neq E_{M}$ (in the case where $E_{m} = E_{M}$ other states maximize the variance as we discuss in \textit{Application 1: Long-range Ising model}).  
The characteristic time scales are derived from \eqref{tau} 
\begin{equation}\label{tauGHZ}
\tau_Z = \frac{1}{x_1 \varepsilon C_{N,k}}\sim \frac{1}{N^k} \ ,\ 
\tau_D  = \frac{4}{x_2 \Lambda^2 C_{N,p} }\sim \frac{1}{N^p} \ ,
\end{equation}
with $\varepsilon=\|H_\mu\|=(\varepsilon_{M})^k-(\varepsilon_{m})^k$, $\Lambda=\Delta L_{\nu}=\lambda_{M}^{(\nu)}-\lambda_{m}^{(\nu)}$, and where $C_{N,m}$ denotes the binomial coefficient $ {N\choose m}$.
This dictates the scaling
\begin{equation}\label{EstErrorGHZ}
\delta x_1 \gtrsim \frac{1}{N^{k-\frac{p}{2}}}\ \ ,\ \ 
\delta x_2 \gtrsim \frac{1}{N^{\frac{p}{2}}}\ ,
\end{equation}
We now discuss the estimation error of the parameter $x_1$. Let us introduce $\delta \equiv k-p/2$. For $k=1=p$ we find that shot noise limit ($\delta=1/2$) \cite{Huelga97}. 
In the presence of decoherence, the optimal case is obtained for $k>p=1$ that gives $\delta=k-1/2$. For all other cases with $k>p\geq 1$ we find that the scaling surpasses the Heisenberg limit as $1 < \delta=k-p/2 < k$.   Notice that for the  time scale $t\ll \tau_D$, the discussion after equations \eqref{EstError} shows that the QCRB scales as $ N^{-\frac{p}{2}}$ (c.f. equation \eqref{tauGHZ}), which differs from the one given after equation (11) in \cite{Alipour14}, i.e.,  $\sim N^{-p}$ in the particular case of a $p$-body Lindblad operators of the form $L_{i_1,\cdots,i_p}=\sigma_{i_1}\otimes\cdots\otimes\sigma_{i_p}$ where $p$ is odd.

\textit{Metrology equivalence of product and GHZ states.}
We next focus on the estimation of the parameter $x_1$. For initial GHZ-states, we have shown that the interrogation time is proportional to the decoherence time $\tau_D$. Nevertheless, this is specific to the maximally-entangled states as the QFI decay time is proportional to $\tau_D$. For initial product state in the context of linear quantum metrology, it is possible 
to compute explicitly the QFI \cite{Huelga97,Alipour14}. In this case, the interrogation time can be related to quantum speed limits, see \cite{delCampo13}. However, this method does not hold in the presence of many-body Hamiltonian or Lindblad operators. In a general case, to compute the decay time of the QFI  which sets  the interrogation time, we propose to use the bounds in equation \eqref{QFIbound1}. 
We assume the initial state to be prepared in a product state $\rho_0 = \bigotimes_{i=1}^{N}\rho_0^{(1)}$ where the initial density matrix of an individual two-level system is chosen to be $\rho_0^{(1)} = |\psi\ra\la\psi|$ with $|\psi\ra=\left(\cos{(\phi)}|\varepsilon_m\ra +\sin{(\phi)}|\varepsilon_M\ra\right)/\sqrt{2}$ where the angle $0<\phi < \pi/2$. Consider a spin chain 
\begin{equation}\label{spinchain}
H_{\text{chain}}=-J\sum_{i_1<\cdots<i_k}\sigma_{i_1}^z\otimes\cdots\otimes\sigma_{i_k}^z\ ,
\end{equation} 
and the master equation
$
\dot{\rho} = -\frac{i}{\hbar}[H_{\text{chain}},\rho]+\mathcal{D}_p(\rho)
$
where we consider ${N\choose p}$ uncorrelated $p$-body dissipators 
\begin{equation}\label{pbodyDissipator}
\mathcal{D}_p(\rho)\equiv \gamma\sum_{i_1<\cdots<i_p}\left(\sigma_{i_1}^z\cdots\sigma_{i_p}^z\rho\sigma_{i_1}^z\cdots\sigma_{i_p}^z-\rho\right)\ .
\end{equation}
Here we restrict the phase to $\phi\neq\pi/4$ as the scaling of the variance of the Hamiltonian \eqref{spinchain} changes from $N^k$ to $N^{k/2}$ for the singular case $\phi=\pi/4$. In \cite{Boixo08} it is shown that the error on the parameter $x_1$ scales as $N^{-\left(k-\frac{1}{2}\right)}$. This scaling is equivalent to the one obtained with a GHZ-state in the presence of decoherence with a $1$-body Lindblad operator. For $p\geq 1$ the effect of the decoherence is not known and requires further technical analysis. Using the bounds in equation \eqref{QFIbound1}, we show that QFI behaves exponentially $\mathfrak{F}_{x_1}\sim N^{-\left(k-\frac{1}{2}\right)}\exp{\left(-\frac{2p}{N\tau_D}t\right)}$, where $\tau_D= \gamma^{-1}p! / N^{p}$. Therefore, the interrogation time is of the order of  $N\times \tau_D \sim N^{-(p-1)}$ which leads to the sensitivity scaling  $N^{-(k-\frac{p}{2})}$. Consequently, we obtain the same scaling for product and GHZ states, using the first bound in equation \eqref{EstErrorGHZ}, which proves their metrology equivalence under many-body decoherence.

\textit{Application 1: Long-range Ising chain.}
So far we have only considered many-body Hamiltonians with site-independent interaction strength, see equations \eqref{Hproduct} and \eqref{spinchain}. However, equations \eqref{kbodyHpbodyLGeneral}-\eqref{EstError} are also valid for site-dependent $k$-body Hamiltonians that can not be written as \eqref{Hproduct}. 
For instance, consider a long-range Ising chain with $p$-body dephasing
\begin{equation}\label{IsingDephasingNS}
\mathcal{L}(\rho)=-\frac{i}{\hbar}[H_I,\rho]+\mathcal{D}_p(\rho)\ ,
\end{equation}
where $H_I = -\sum_{i<j}J_{ij}\sigma_i^z\sigma_j^z$, with $J_{ij}=J/|i-j|^\alpha$, where the exponent $\alpha\geq 0$ controls the range of the interactions, and where the $p$-body dephasing $\mathcal{D}_p(\rho)$ is defined in equation \eqref{pbodyDissipator}. 
The experimental realization of this model without dissipation has recently been reported \cite{Richerme14,Jurcevic14}.
The dephasing could be obtained via a site-dependent classical noise \cite{CBCdC16}.   
We investigate the effect of the dephasing and of the range $\alpha$ on the estimation error of the parameter $J$. 
We introduce the dimensionless Hamiltonian $H\equiv H_I/J$ and give a scaling of the left hand side bound in equation \eqref{EstError} (setting $x_1=J$) by computing the variance of the Hamiltonian $H$, see equation \eqref{tau}.  The seminorm of the Hamiltonian $\|H\| = E_M-E_m = 2\sum_{i\leq\lfloor \frac{N}{2}\rfloor}\sum_{\lfloor \frac{N}{2}\rfloor < j}(j-i)^{-\alpha}$ maximizes its variance, see \cite{Boixo07}. After using the Euler-Maclaurin formula we find the scaling for the variance of $H$ for large $N$, as detailed in \cite{SM},
$
\Delta H \sim  N^{2-\alpha} 
$ 
for $\alpha<2$, $\log(N)$ for $\alpha=2$ and $1$ for $\alpha>1$. 
Therefore, $\tau_Z \sim N^{-\beta}$ with $\beta>1/2$ only for $\alpha<3/2$. We also show that the sum of the variances of the Lindblad operators $\sum_i (\Delta\sigma_i)^2$ scales as $\frac{N}{2}$. Thus, it follows from our previous results, see equations \eqref{tau} and \eqref{EstError},  that the scaling of  the estimation error for the parameter $J$ is set by the range of the interactions according to $N^{-\delta}$ with $\delta\equiv 2-\alpha-\frac{p}{2}$ which beats the shot-noise limit for $\alpha < \frac{3}{2}-\frac{p}{2}$, gives the Heisenberg and super-Heisenberg limit for $\alpha\leq 1-\frac{p}{2}$, and leads to a maximum precision with exponent $2-\frac{p}{2}$ for $\alpha=0$. For instance, for $p=1$ super-Heisenberg limit is robust for $\alpha < 1/2$ while for $p=2$ we have $\frac{1}{2}<\delta\leq 1$ for $0\leq\alpha<\frac{1}{2}$. Interestingly, for $\alpha<3/2$ we can demonstrate the metrology equivalence between the states maximizing the variance of the Hamiltonian and product states of the form above equation \eqref{spinchain} for $\phi\neq \pi/4$, see \cite{SM} for further details. 

\textit{Application 2: Fluctuating many-body Hamiltonians.}
In experimental platforms, a parameter $x_1$ often exhibits a stochastic component 
$
x_1\mapsto x_1 + \hbar\sqrt{\gamma}\eta(t)\ , 
$
where $\eta(t)$ is a real Gaussian noise with amplitude $\gamma$. To optimize control on the system it is necessary to have an accurate estimate of the amplitude of the noise $\gamma$, that can be obtained via  the method we have developed.  Consider the spin chain Hamiltonian given in equation \eqref{spinchain} where we add a white noise to the parameter $J$. The density matrix $\rho=\la \rho_{st}\ra$ obtained by averaging over stochastic realizations fulfills the  master equation \cite{Milburn91,Budini01,CBCdC16}
$$ 
\frac{d}{dt}\rho = -\frac{i}{\hbar}[H_{\text{chain}},\rho] + \gamma \left(L\rho L - \frac{1}{2}L^2\rho-\frac{1}{2}\rho L^2\right)\ ,
$$   
where the $k$-body Lindblad operator is symmetrized $L = \sum_{i_1<\cdots < i_k}\sigma_{i_1}^z\otimes\cdots\otimes\sigma_{i_k}^z$, and the $k$-body Hamiltonian is given by equation \eqref{spinchain}. We assume that the initial state maximizes the variance of the Hamiltonian $H_{\text{chain}}$ (e.g., a GHZ-state for $k$ odd). A part from being non-local, the resulting dissipator is a symmetrized sum of Lindblad operators. The decoherence it induces \cite{CBCdC16} is sped-up by the correlation between many-body Lindblad operators $\tau_D\sim N^{-2k}$. As a result, as shown by the second bound in equation \eqref{EstError}, the precision of the estimation of the parameter $\gamma$ is significantly enhanced
$ 
\delta\gamma \sim N^{-k}\ . 
$
This is to be contrasted with the scaling $N^{-\frac{k}{2}}$ associated with an uncorrelated dissipator describing the decoherence of each particle by an independent bath, see master equation \eqref{MasterEq2}. 

\textit{Conclusion and outlook.}
We have introduced a method to study the precision of the parameter estimation in nonlinear quantum metrology of many-body open quantum systems in the Markovian limit. We have exploited it to determine the achievable precision when the dephasing dynamics is generated by a $k$-body Hamiltonian and dissipators with $p$-body Lindblad operators. We have shown that for initial GHZ-states the scaling of the estimation error of the Hamiltonian parameter $x_1$ scales as $ N^{-\left(k-\frac{p}{2}\right)}$. Consequently, the precision of the estimation beats the shot-noise limit when $2k-1 > p\geq  1$. The metrology equivalence of product and GHZ state states has been demonstrated. Our method can also be applied to many-body systems for which the coupling constants depend on the distance between sites, as we have demonstrated in the long-range Ising model. In the latter case,  the range of the interaction affects drastically the precision of the Hamiltonian parameter estimation. In addition, we have shown that the precision of the estimated system-environment coupling $x_2$ is limited by $N^{-\frac{p}{2}}$, while for fluctuating $k$-body Hamiltonian systems an enhanced scaling $N^{-k}$ governs the  precision of  the amplitude of the noise. Our results have broad application in quantum metrology of many-body open quantum systems \cite{Barreiro11} and certification of quantum simulations. 
We hope this article will motivate future investigations on nonlinear quantum metrology in the presence of dissipation and for non-markovian dynamics \cite{Chin12}.

{\it Acknowlegments.---}  We acknowledge funding support by UMass Boston (project P20150000029279) and the John Templeton Foundation. MB thanks the French Revolution for introducing the International System of Units (SI).

\newpage 

\appendix

\begin{widetext}

\section{QFI for $x_1$: New bounds}\label{SM:QFIbound1}

In the main body of the paper, we use the following expression for the QFI  (see equation (4))
\begin{equation}\label{SM:QFI1}
\mathfrak{F}_{x_1}(t)=\sum_{n,n':\xi_n+\xi_{n'}\neq 0}4\xi_n \frac{|\la \xi_n| \partial_{x_1}\rho(t)|\xi_{n'}\ra|^2}{(\xi_n+\xi_{n'})^2} \ ,
\end{equation}
where $|\xi_n\ra$ are the eigenstates of the density matrix $\rho$. For a Markovian equation of the form (see also equation (1) in the main body)
\begin{equation}\label{SM:MasterEq}
 \frac{d}{dt}\rho(t)=\mathcal{L}(\rho(t)) = -\frac{ix_1}{\hbar}[H,\rho(t)] + \sum_{\alpha}x_\alpha\left(L_\alpha\rho(t) L_\alpha -\frac{1}{2}\left\{L_\alpha^2,\rho(t)\right\}\right)\ , 
\end{equation} 
there exists a quantum semigroup \cite{Lindblad76,Breuer02} so that 
$$\rho(t) = \exp\left(t\mathcal{L}\right)\rho(0)\ .$$
Hence it follows that
$$ \partial_{x_1}\rho(t) = -\frac{it}{\hbar}[H,\rho(t)]\ .$$
Hence, one can rewrite \eqref{SM:QFI1} as
\begin{equation}\label{SM:QFI2}
\mathfrak{F}_{x_1}(t)=\frac{4t^2}{\hbar^2}\sum_{n,n':\xi_n+\xi_{n'}\neq 0}\xi_n \frac{|\la \xi_n| [H,\rho]|\xi_{n'}\ra|^2}{(\xi_n+\xi_{n'})^2} \ ,
\end{equation}
with 
$$ 
\la\xi_n| [H,\rho]|\xi_{n'}\ra = \la\xi_n| [H-\la H\ra_\rho,\rho]|\xi_{n'}\ra = \left(\xi_{n'}-\xi_{n}\right)\la\xi_n| (H-\la H\ra_\rho)|\xi_{n'}\ra\ .
$$
Inserting the last equation into \eqref{SM:QFI2}, we find
\begin{equation}\label{SM:QFI3}
\mathfrak{F}_{x_1}(t)=\frac{4t^2}{\hbar^2}\sum_{n,n':\xi_n+\xi_{n'}\neq 0}\xi_n \frac{(\xi_n-\xi_{n'})^2}{(\xi_n+\xi_{n'})^2}\left|\la\xi_n| (H-\la H\ra_\rho)|\xi_{n'}\ra\right|^2  \ .
\end{equation}
We introduce two bounds
\begin{equation*}
c_m \leq\frac{(\xi_n-\xi_{n'})^2}{(\xi_n+\xi_{n'})^2}\leq c_M  \ ,
\end{equation*}
with 
$$
c_{m}\equiv \text{min}_{k\neq p}\left(\frac{\xi_k-\xi_p}{\xi_k+\xi_p}\right)^2\ ,\ c_{M}\equiv \text{max}_{k\neq p}\left(\frac{\xi_k-\xi_p}{\xi_k+\xi_p}\right)^2\ .
$$
The function $x\mapsto \frac{(1-x)^2}{(1+x)^2}$ is a decreasing function of $x$ in the interval $[0,1]$, therefore
\begin{equation}\label{SM:Coef}
c_{m}\equiv \left(\frac{1-r}{1+r}\right)^2\ ,\ c_{M}\equiv \left(\frac{\xi_{M}-\xi_{m}}{\xi_{M}+\xi_{m}}\right)^2\ ,
\end{equation}
where $\xi_{M}\geq  \cdots\geq \xi_{m}$, i.e., $\xi_{M(m)}$ is the maximum (minimum) eigenvalue of $\rho$ and $r=\max\limits_{n<n'}\left(\frac{\xi_n}{\xi_{n'}}\right)$. Here, we have made use of the fact that the minimum value of the ratio of eigenvalues $\min\limits_{n<n'}\left(\frac{\xi_n}{\xi_{n'}}\right)$ equals $\lambda_m/\lambda_M$.

Summing over the index $n'$ in equation \eqref{SM:QFI3} yields
$$
\sum_{n'}\left|\la\xi_n| (H-\la H\ra_\rho)|\xi_{n'}\ra\right|^2 = \la\xi_n| (H-\la H\ra_\rho)^2|\xi_{n}\ra\ , 
$$
and after summing over $n$ we obtain 
$$
\sum_n \xi_n \sum_{n'}\left|\la\xi_n| [H-\la H\ra_\rho,\rho]|\xi_{n'}\ra\right|^2 = \sum_n \xi_n\la\xi_n| (H-\la H\ra_\rho)^2|\xi_{n}\ra = \la(H-\la H\ra_\rho)^2\ra_\rho =  \Delta_{\rho} H^2\ ,
$$
where $\Delta_{\rho} H^2 = \la H^2 \ra_{\rho} - \la H \ra_{\rho}^2$ is the variance of the Hamiltonian in the state $\rho$. Gathering the last equations we find 
\begin{equation}\label{SM:QFIBound} 
c_m \frac{4t^2\Delta_{\rho} H^2}{\hbar^2} \leq\mathfrak{F}_{x_1}(t) \leq c_M \frac{4t^2\Delta_{\rho} H^2}{\hbar^2}\ ,
\end{equation}
which gives equations (5)-(6) in the main body. 

\section{Derivation of the QFI for a maximum variance state}\label{SM:QFImax}

The initial state is $\rho_0=|\Psi_0\ra\la \Psi_0 |$ where $|\Psi_0\ra = \left(|E_m\ra+|E_M\ra\right)/\sqrt{2}$ and where $|E_{m(M)}\ra$ is the eigenstate of $H$ associated with the minimum (maximum) eigenvalue $E_{m(M)}$. The computational basis is given by the two eigenstates 
$$|e_0\ra \equiv |E_m\ra = \bmat 1 \\ 0 \emat,\ |e_0\ra \equiv |E_M\ra = \bmat 0 \\ 1 \emat\ .$$
The time evolution of the initial density matrix 
$$\rho_0 = \frac{1}{2}\bmat 1 & 1 \\ 1 & 1 \emat\ ,$$
under the master equation (7) in the main body is given by   
$$\rho(t) = \frac{1}{2}\bmat 1 & e^{\alpha} \\ e^{\alpha^\ast} & 1 \emat\ ,$$
where 
$$ \alpha = \frac{ix_1}{\hbar}\epsilon t + \frac{x_2}{2}\lambda^2 t\ , $$
with $\epsilon = E_M-E_m $ and $\lambda^2 = \sum_{\nu} \left(\lambda_{M}^{(\nu)}-\lambda_{m}^{(\nu)}\right)^2$. Here,  $\lambda_{m(M)}^{(\nu)}$ denotes the eigenvalue of $L_\nu$ associated with the state $|E_{m(M)}\ra$.
It is easy to check that $\epsilon = 2\Delta_{\rho_0} H$ and $\lambda^2 = 4\sum_{\nu}\Delta_{\rho_0} L_\nu^2$. Hence, using the  definitions given in equation (10) in the main body of the article, we can rewrite the rate $\alpha$ as a function of both the Zeno and the decoherence times 
$$ \alpha = -i\frac{t}{\tau_Z} - \frac{2t}{\tau_D}\ , $$
where 
$$
\tau_Z = \frac{\hbar}{2x_1\Delta_{\rho_0} H} = \frac{\hbar}{x_1\epsilon}\ ,\ \ \tau_D = \frac{1}{x_2 \sum_{\nu}\Delta_{\rho_0} L_\nu^2} = \frac{4}{x_2 \lambda^2}\ .
$$
The eigenvalues of the density matrix read
$$
\xi_{\pm} = \frac{1 \pm e^{2\mathrm{Re}(\alpha)}}{2} = \frac{1 \pm e^{-2t/\tau_D}}{2}\ ,
$$
and the associated eigenvectors are
$$
| \xi_{\pm} \ra = \frac{1}{\sqrt{2}}\bmat \pm e^{\mathrm{Im}(\alpha)} \\ 1 \emat = \frac{1}{\sqrt{2}}\bmat \pm e^{-i t/\tau_Z} \\ 1 \emat \ .
$$
From the previous results we can compute explicitly the purity
$$p(t)\equiv\text{tr}\left(\rho(t)^2\right) = \xi_{+}^2 + \xi_{-}^2 = \frac{1+e^{-4t/\tau_Z}}{2}\ ,$$
as well as the quantum fidelity
$$F(t)\equiv\text{tr}\left(\rho(t)\rho_0\right)=\text{tr}\left[\frac{1}{4}\bmat 1+e^{\alpha} & 1+e^{\alpha} \\ 1+e^{\alpha^\ast} & 1+e^{\alpha^\ast} \emat\right] = \frac{1}{2}\left(1+\mathrm{Re}\left(e^\alpha\right)\right) 
= \frac{1}{2}\left(1 + e^{-2 t/\tau_D}\cos{\left(t/\tau_Z\right)}\right) \ ,$$
consistently with the equation (9) in the main body of the letter. Notice that the purity is a monotonically decreasing function of $t$ consistently with the fact that the Lindbladian in equation (7) in the main body of the letter is unital \cite{Lidar06}.

Next, we compute the QFI using equation \eqref{SM:QFI1} and the above results.  
First of all, let us note that
\begin{align*}
&\partial_{x_1} \rho(t) |\xi_{\pm}\ra = \frac{1}{2\sqrt{2}} \bmat 0 & -i\sigma e^{-i\sigma x_1} e^{-\beta x_2} \\ i\sigma e^{i\sigma x_1} e^{-\beta x_2} & 0 \emat \cdot \bmat \pm e^{-i\sigma x_1} \\ 1 \emat = \frac{i\sigma}{2\sqrt{2}}e^{-\beta x_2}\bmat \mp e^{-i\sigma x_1} \\ 1 \emat = \frac{\pm i\sigma}{2\sqrt{2}}e^{-\beta x_2}|\xi_{\mp}\ra\ , \\
&\partial_{x_2} \rho(t) |\xi_{\pm}\ra = \frac{1}{2\sqrt{2}} \bmat 0 & -\beta e^{-i\sigma x_1} e^{-\beta x_2} \\ -\beta e^{i\sigma x_1} e^{-\beta x_2} & 0 \emat \cdot \bmat \pm e^{-i\sigma x_1} \\ 1 \emat = \frac{\mp\beta}{2\sqrt{2}}e^{-\beta x_2}\bmat \pm e^{-i\sigma x_1} \\ 1 \emat = \frac{\mp\beta}{2\sqrt{2}}e^{-\beta x_2}|\xi_{\pm}\ra\ ,
\end{align*}
where $\sigma = \epsilon t/\hbar = t/(x_1\tau_Z)$, and $\beta=\frac{1}{2}\lambda^2 t = 2t/(x_2\tau_D)$.
Hence, we find
\begin{subequations}\label{SM:Eigenvalues}
\begin{equation}
\la \xi_{\pm}|\partial_{x_1} \rho(t) |\xi_{\pm}\ra = 0,\ \text{and}\ \left|\la \xi_{\pm}|\partial_{x_1} \rho(t) |\xi_{\mp}\ra\right|^2 = \frac{\sigma^2}{8}e^{-2\beta x_2}\ , 
\end{equation}
\begin{equation}
\left|\la \xi_{\pm}|\partial_{x_2} \rho(t) |\xi_{\pm}\ra\right|^2 =  \frac{\beta^2}{8}e^{-2\beta x_2},\ \text{and}\ \la \xi_{\pm}|\partial_{x_2} \rho(t) |\xi_{\mp}\ra = 0\ , 
\end{equation}
\end{subequations}
From the previous equations and using \eqref{SM:QFI1}, we obtain 
\begin{align*}
&\mathfrak{F}_{x_1}(t) = \sigma^2 e^{-2\beta x_2} = \frac{t^2}{x_1^2\tau_Z^2} e^{- 4t/\tau_D}\ , \\
&\mathfrak{F}_{x_2}(t) = \beta^2 \frac{e^{-2\beta x_2}}{1-e^{-2\beta x_2}} = \frac{t^2}{x_2^2\tau_D^2} \frac{e^{- 4 t/\tau_D}}{1-e^{- 4 t/\tau_D}}\ , 
\end{align*}
as $\xi_{+}+\xi_{-}=1$ and $\xi_{+}^{-1}+\xi_{-}^{-1}=4(1-e^{-2\beta})^{-1}$. \\
Notice that $\mathfrak{F}_{x_1}(t)$ could have been obtained using the bounds given by equation \eqref{SM:Coef} in equation \eqref{SM:QFIBound} as
$$ c_M = c_m = \frac{\xi_{+}-\xi_{-}}{\xi_{+}+\xi_{-}} = e^{- 4t/\tau_D} $$ and $\Delta_{\rho_0}H=\hbar/(2\tau_Z)$.
% by equation \eqref{tau}.

To estimate the precision, we use the quantum Cramer-Rao bound (QCRB), see equation (2) in the main body of this article. We choose the interrogation time to be $t_{i}=\mu_i \tau_D$, where $\mu_i$ is such that $d\mathfrak{F}_{x_i}(t)/dt=0$. The repetition number are $\nu_i=T/t_i,\ i=1,2$, where $T$ is the total time of the metrology protocol.  
For the parameter $x_1$ we find $\mu_1=1/2$ to be the unique solution of the equation 
$ \frac{d}{d\mu}\left(\mu^2 e^{-4\mu}\right)=0\ .  $
For the second parameter $x_2$, the solution of the equation
$\frac{d}{d\mu}\left(\mu^2 e^{-4\mu}(1-e^{-4\mu})^{-1}\right)=0$ is $\mu_2\approx 0.40(1)$. 
Substituting the interrogation time and the repetition number in the QFIs above we find equation (12) in the paper, with $\kappa_1 = \mu_1^{1/2}e^{-2\mu_1}$ and $\kappa_2 = \mu_2^{1/2}\left(e^{4\mu_2}-1\right)^{-1/2}$.\\

\textit{Application to GHZ-states.} Consider that the $k$-body Hamiltonian as the form of a product 
$$ 
H_{\mu} = H_{i_{1}}\otimes\cdots\otimes H_{i_{k}}\ , 
$$
where $\mu=(i_1,\cdots,i_k)$ with $1\leq i_1<i_2<\cdots <i_k\leq N$, 
and that the Hamiltonian is symmetrized 
$$
H=\sum_{\mu}H_{\mu} = \sum_{1\leq i_1<\cdots <i_k\leq N}H_{i_{1}}\otimes\cdots\otimes H_{i_{k}}\ .
$$
If the initial state is a GHZ-state, we have
$$
|\Psi_0\ra = \frac{|\varepsilon_m\ra^{\otimes N}+|\varepsilon_M\ra^{\otimes N}}{\sqrt{2}}\ ,
$$
where $\varepsilon_{m(M)}$ is the minimal (maximal) eigenvalue of the single-body Hamiltonian $H_{i}$. In this case we have 
$$
\epsilon = E_M-E_m = {N\choose k}\times\|H_\mu\|={N\choose k}\times\left[(\varepsilon_{M})^k-(\varepsilon_{m})^k\right]\ ,
$$ 
and 
$$\lambda^2 = \sum_{\nu} \left(\lambda_{M}^{(\nu)}-\lambda_{m}^{(\nu)}\right)^2 = {N\choose p}\times \left(\lambda_{M}^{(\nu)}-\lambda_{m}^{(\nu)}\right)^2\ .
$$ 
Hence we find that
$$
\tau_Z= \frac{\hbar}{x_1\epsilon} = \frac{\hbar}{x_1 {N\choose k}\times \varepsilon}\ ,\ \ \tau_D = \frac{1}{x_2 \lambda^2} = \frac{1}{x_2 {N\choose p}\times \Lambda^2}\ ,
$$
where $\varepsilon \equiv (\varepsilon_{M})^k-(\varepsilon_{m})^k$ denotes the seminorm of the $k$-body operator $H_{\mu}$, and $\Lambda \equiv \left(\lambda_{M}^{(\nu)}-\lambda_{m}^{(\nu)}\right)^2$ which is independent of $\nu$.  
For large $N\gg k$ we have the asymptotic ${N\choose k}\sim \frac{N^k}{k!}$, showing equation (14) in article. 

To conclude this part, we point out that the joint estimation strategy does not modify the precisions of the estimations. This is due to the fact that the dephasing only changes the eigenvalues of the density matrix while the Hamiltonian rotates the eigenvectors as one can see in equation \eqref{SM:Eigenvalues}. This implies that the off-diagonal term of the QFI (in the matrix form, see \cite{Helstrom76,Holevo82,Humphreys13,Cheng14}) vanishes
$$
\mathfrak{F}_{12} = \sum_{k,l}\frac{2\langle \xi_k|\partial_{x_1}\rho|\xi_l\rangle\langle \xi_l|\partial_{x_2}\rho|\xi_k\rangle}{\xi_{k}+\xi_{l}}\ ,
$$
as $\langle \xi_k | \partial_{x_1}\rho | \xi_l \rangle= 0$ if $k = l$, $\langle \xi_k | \partial_{x_2}\rho | \xi_l\rangle = 0$ if $k\neq l$. 

\section{$N$ two-level systems: orthonormal basis}\label{SM:SpinONB}

In the following sections we will need to compute the eigenvalues of Linbdlad operators and Hamiltonians for two-level systems. In order to use consistent notations and definition, we devote this section to set up an  orthonormal basis (ONB) for $N$ two-level systems. This is applicable for $N$ spins systems where the spin up corresponds to the highest energy level and the spin down to the lowest energy level. 

Consider $N$ particles with two levels of energy $|\pm\ra\in\mathcal{H}$, where $\mathcal{H}$ is the single-particle Hilbert space. A $N$-particle state $|\Psi\ra\in \mathcal{H}^{\otimes N}$ is a vector in the Hilbert space $\mathcal{H}^{\otimes N}$. The corresponding density matrix $\rho$ has a matrix representation in the $N$-particle basis of $2^N$ combinations of upper and lower states $|a_1 a_2\cdots a_N \ra = \bigotimes_{j=1}^{N}|a_j\ra $, where the kets $|a_j\ra=|+\ra$ or $|-\ra$ represent the eigenstates of a single particle Hamiltonian $H_j$ so that $H_j|-(+)\ra = \varepsilon_{m(M)}|-(+)\ra$. For $N$ spins systems, this Hamiltonian is equal to the single Pauli matrices $\sigma_j^z$, i.e., $\sigma_j^z|\pm\ra =\pm 1|\pm\ra$.

Let us construct the computational basis. We first set $N+1$ vectors
$$
|v_k\ra=|\underbrace{-1\cdots -1}_{N-k\ \text{times}}\underbrace{+1\cdots +1}_{k\ \text{times}} \ra\ ,\ k=0,\cdots,N\ ,
$$
and construct all the vector in the basis as follows
$$|e_k^{(m)}\ra=\pi^{(m)}(|v_k\ra)\ ,$$
where the $\pi^{(m)}$'s are permutations in the $S_{N}$ symmetric group, and where the label $m$ goes from $1$ to ${N \choose k}$. By convention $\pi^{(1)}$ denotes the identity so that $|e_k^{(1)}\ra = |v_k\ra$. After summing over all possible $k$ and $m$ we obtain the correct number of $2^N=\sum_{k=0}^{N}{N \choose k}$ orthonormal vectors in the basis.
As a shorthand, we denote the $2^N$ vector in the ONB $|e_j\ra,\ j=1,2,3,\cdots,2^N$, satisfying the identity
$$\mathbb{I}=\sum_j |e_j\ra\la e_j| = \sum_{k=0}^N \sum_{m}|e_k^{(m)}\ra\la e_k^{(m)}|\ , $$  
where $\mathbb{I}$ is the identity operator in the Hilbert space $\mathcal{H}^{\otimes N}$. \\ 

\fbox{\begin{minipage}{50em}
\textit{Example 1: orthonormal basis for $N=4$} \\
We first find the $4+1=5$ vectors
$$|v_0\ra =| - - - - \ra\ ;\ |v_1\ra =| - - - + \ra\ ;\ |v_2\ra =| - - + + \ra;\ |v_3\ra =| - + + + \ra;\ |v_4\ra =| + + + + \ra\ ,$$
that define the first $5$ vectors of the basis $|e_k^{(1)}\ra = |v_k\ra,\ k=0,1,2,3,4$.
Then, we construct the other $2^4-5=16-5=11$ vectors by permutations
$$|e_1^{(2)}\ra =| - - + - \ra\ ;\ |e_1^{(3)}\ra=| - + - - \ra\ ;\ |e_1^{(4)}\ra=| + - - - \ra\ ,$$
$$|e_2^{(2)}\ra =| - + - + \ra\ ;\ |e_2^{(3)}\ra=| + - - + \ra\ ;\ |e_2^{(4)}\ra=| - + + - \ra\ ;\ |e_2^{(5)}\ra=| + - + - \ra\ ;\ |e_2^{(6)}\ra=| + + - - \ra\ ,$$
$$|e_3^{(2)}\ra =| + + - + \ra\ ;\ |e_3^{(3)}\ra=| + - + + \ra\ ;\ |e_3^{(4)}\ra=| - + + + \ra\ .$$
From above, we find $1$, $4$, $6$, $4$, $1$ states for $k=0,\ 1,\ 2,\ 3,\ 4$ respectively, which in total gives $1+4+6+4+1=16=2^4$ states.
\end{minipage}}
%\linebreak\\ \linebreak\\ 

\section{Eigenvalues of $k$-body spin operators}\label{SM:evSpin}

Following the construction of the $N$-spin ONB, we are now interested in computing the eigenvalues of $k$-body spin operators
\begin{equation}
S_{\nu}=\sigma_{i_1}^z\cdots\sigma_{i_k}^z\ ,\ \nu=\{1\leq i_1<\cdots<i_k\leq N\}\ .
\end{equation}

\textit{Case $k=2$.}
To compute the eigenvalues explicitly, it suffices to notice that
\begin{equation} \label{SM:Spectsigmaij}
\sigma_i^z\sigma_j^z | e_q^{(1)}\ra = \left\{ \begin{array}{ll} +1,\ \text{if}\ i \ \text{and}\ j\leq\ \text{or}\ \geq N-q\ , \\ -1,\ \text{otherwise}\ .\end{array} \right.
\end{equation}
and to compute the number of terms in the expression of $L$ in each case
\begin{subequations}\label{k=2Degeneracy}
\begin{equation}\label{k=2Degeneracy+1}
\text{Number}\left(\sigma_i^z\sigma_j^z\ \text{with}\ i \ \text{and}\ j\leq\ \text{or}\ \geq N-q\right)={N-q \choose 2}+{q \choose 2} = \frac{1}{2}\Big((N-q)(N-q-1)+q(q-1)\Big)\ ,
\end{equation}
\begin{equation}\label{k=2Degeneracy-1}
\text{Number}\left(\sigma_i^z\sigma_j^z\ \text{with}\ i \ \text{or}\ j\geq N-q\right)={N \choose 2}-{N-q \choose 2}-{q \choose 2} = {q \choose 1}\times {N-q\choose 1} = q\left(N-q\right)\ .
\end{equation}
\end{subequations}
This yields the expression of the degeneracy of the eigenvalues $\pm 1$, see equations \eqref{k=2Degeneracy} . \\

\textit{Case $k\geq 2$.} 
We generalize the previous particular case to $k\geq 2$. We consider $k$ odd, the generalization to $k$ even is straightforward. We also assume that $N\gg k$. First, we note that
\begin{equation} \label{SM:SpectSN}
S_{\nu} | e_q^{(1)}\ra = \left\{ \begin{array}{ll} 
-1,\ \text{if}\ i_1<\cdots<i_{k} \leq N-q\ , \\
+1,\ \text{if}\ i_1<\cdots<i_{k-1} \leq N-q,\ \text{and}\ i_k>N-q\ , \\
\vdots \\ 
-1,\ \text{if}\ i_1\leq N-q,\ \text{and}\ N-q<i_2<\cdots<i_k\ , \\
+1,\ \text{if}\ N-q<i_1<\cdots<i_k\ .\end{array} \right.
\end{equation}
The number of terms in the expression of $L$ in each case is 
\begin{align*}
\text{Number}\left(\sigma_{i_1}^z\cdots\sigma_{i_k}^z\ \text{with}\ i_1<\cdots<i_{k} \leq N-q\right)&=1\ ,\\
\text{Number}\left(\sigma_{i_1}^z\cdots\sigma_{i_k}^z\ \text{with}\ i_1<\cdots<i_{k-1} \leq N-q,\ \text{and}\ i_k>N-q\right)&={q\choose 1}\times {N-q\choose k-1}\ ,\\
&\vdots \\ 
\text{Number}\left(\sigma_{i_1}^z\cdots\sigma_{i_k}^z\ \text{with}\ i_1<\cdots<i_{r} \leq N-q,\ \text{and}\ N-1<i_{r+1}<\cdots<i_k\right)&={q\choose r}\times {N-q\choose k-r}\ ,\\
&\vdots \\ 
\text{Number}\left(\sigma_{i_1}^z\cdots\sigma_{i_k}^z\ \text{with}\ N-q<i_1<\cdots<i_{k}\right)&={q\choose k}\ ,
\end{align*}
where we conventionally set ${n\choose m}=0$ if $n<m$. We find the degeneracy of the eigenvalues $\pm 1$ 
\begin{subequations}\label{Degk}
\begin{equation}\label{Degk+1}
\ \ \ \ \ \ \ \ \ \ \ \ \ \ \ \ \ \ \ \text{Deg}\left(\zeta = +1\right) = \sum_{s=0}^{(k-1)/2}{q\choose 2s+1}{N-q\choose k-(2s+1)}\ , 
\end{equation}
\begin{equation}\label{Degk-1}
\text{Deg}\left(\zeta = -1\right) = \sum_{s=0}^{\lfloor k/2\rfloor}{q\choose 2s}{N-q\choose k-2s}\ .
\end{equation}
\end{subequations}
After summing over the degeneracy in equations \eqref{Degk} we find via the  Chu-Vandermonde identity that the total degeneracy is equal to 
$$
\sum_{s=0}^{k}{q\choose s}{N-q\choose k-s} = {N\choose k}
$$
consistently with the total number of $k$-body spin operators.  

Similarly, for $k$ even we obtain that equations \eqref{Degk+1} and \eqref{Degk-1} gives the degeneracy of the eigenvalue $-1$ and $+1$ respectively. 

\section{Initial product-state}\label{SM:Matrix:ProductState}

In this section we want to estimate the constants $c_M$ and $c_m$, see equation \eqref{SM:Coef} and equations (5)-(6) in the paper, for a $p$-body dissipator of the form
\begin{equation}\label{SM:pbodyDephasing}
\mathcal{D}_p(\rho)=-\gamma\sum_{i_1<\cdots<i_p}\left(\sigma^z_{i_1}\cdots\sigma^z_{i_p}\ \rho\ \sigma^z_{i_1}\cdots\sigma^z_{i_p}-\rho\right)\ .
\end{equation}
We consider the pure dephasing channel \eqref{SM:pbodyDephasing} without the $k-$body Hamiltonian $H$ as we assume it commutes with the $p$-body Lindblad operator $L_\nu=\sigma^z_{i_1}\cdots\sigma^z_{i_p},\ \nu=\{i_1<\cdots<i_p\}$, and hence preserves the spectrum. In other words, we can absorb the Hamiltonian part in the interaction picture which makes the results below general under the assumption $[H,L_\nu]=0$.\\

For product states of the form $\rho=\bigotimes_{i=1}^N|\phi_0\ra\la\phi_0|$ with $|\phi_0\ra = (|+\ra+|-\ra)/\sqrt{2}$, the solution of the master equation is
$\rho_{ij}(t)=\rho_{ij}(0)e^{-\frac{\gamma}{2}\lambda_{ij}^2 t} =\frac{1}{2^N}e^{-\frac{\gamma}{2}\lambda_{ij}^2 t} $ as the initial density matrix has the matrix elements given by $\rho_{ij}(0)= 2^{-N}$ in the ONB $\{|e_{j}\ra\}_{j=1}^{2^N}$.
The rates are given by $\lambda_{ij}^2=\sum_{\nu\in\mathfrak{S}_p}\left(\lambda_i^{(\nu)}-\lambda_j^{(\nu)}\right)^2$.  
The argument below is rotation-invariant and is valid for any initial states of this form
$$
\rho_0^{(\phi)} = \left( \cos{(\phi)}| +\ra +  \sin{(\phi)}| -\ra \right)^{\otimes N},\  0 < \phi < \frac{\pi}{2}\ ,
$$
as $\rho_0^{(\phi)} = R_{\phi-\pi/4} \rho_0^{(\pi/4)}R_{\phi-\pi/4}^\dagger $ with $R_{\phi-\pi/4} = r_{\phi-\pi/4}^{\otimes N}$ where $\cos{(\phi)}| +\ra +  \sin{(\phi)}| -\ra =r_{\phi-\pi/4}\left(\frac{|+\ra+|-\ra}{\sqrt{2}}\right) $ and as the rotation operator $R_{\phi-\pi/4} $ leaves invariant the characteristic equation of $\rho_t$. 

We first consider  the maximal eigenvalue state. This state is clearly the initial state $|\Psi_0\ra = \frac{1}{\sqrt{2^N}}\sum_{i}|e_i\ra\equiv |\xi_M\ra$,  
\begin{align}
\rho(t)|\Psi_0\ra = \frac{1}{\sqrt{2^N}}\sum_{i,j,l}\rho_{ij}(t)|e_i\ra\la e_j|e_l\ra = \frac{1}{\sqrt{2^N}}\sum_i\left(\sum_j \rho_{ij}(t)\right)|e_j\ra \ .
\end{align}
By symmetry, we show that the sum over all the elements in one column $\sum_j \rho_{ij}(t)$ is invariant, which means that  
\begin{align}
\rho(t)|\Psi_0\ra = \xi_M(t)|\Psi_0\ra\ ,\ \text{with}\ \xi_M(t)=\frac{1}{2^N}\left(1+\sum_{\alpha}\kappa_{M}^{(\alpha)}e^{-\frac{\gamma}{2}\Lambda_{\alpha}^2 t}\right)\ ,
\end{align}
where $\Lambda_{\alpha}$ denotes the different values of $\lambda_{ij}$ and the coefficient $\kappa_{M}^{(\alpha)}$ equals the degeneracy of these values.

Any other eigenstate $|\xi_n\ra$ follows this decomposition
$$
|\xi_n\ra = \sum_{i}a_i^{(n)}|e_i\ra \ , 
$$
which leads to
\begin{align}
\rho(t)|\xi_n\ra = \sum_{i,j,l}\rho_{ij}(t)|e_i\ra\la e_j|a_l^{(n)}|e_l\ra = \sum_i\left(\sum_j \rho_{ij}(t)a_j^{(n)}\right)|e_i\ra \ .
\end{align}
As a result,  we find the equation
\begin{equation}
\sum_{ij}\rho_{ij}(t)a_{j}^{(n)} = \xi_n a_i^{(n)} \Leftrightarrow \frac{1}{2^N}\left(a_i^{(n)}+\sum_{j\neq i} e^{-\frac{\gamma}{2}\lambda_{ij}^2 t}a_{j}^{(n)}\right) = \xi_n a_i^{(n)}\ ,\ \forall i\ \text{with}\ a_{i}\neq 0\ ,
\end{equation}
that implies
\begin{equation}
\xi_n =\frac{1}{2^N}\left(1+\sum_{j\neq i}\frac{a_{j}^{(n)}}{a_i^{(n)}} e^{-\frac{\gamma}{2}\lambda_{ij}^2 t}\right)\ ,\ \forall i\ \text{with}\ a_{i}\neq 0\ ,
\end{equation} 
Therefore,  we obtain the general form
\begin{equation}\label{evrho}
\xi_n =\frac{1}{2^N}\left(1+\sum_{\alpha}\kappa_{n}^{(\alpha)} e^{-\frac{\gamma}{2}\Lambda_{\alpha}^2 t}\right)\ ,
\end{equation} 
with the conditions $\sum_{n}\kappa_{n}^{(\alpha)} = 0$ imposed by $\sum_n\xi_n=1$ for all $t$.
In what follows, we do not consider the zero-eigenvalue because the corresponding eigenvectors are also eigenvectors of the Lindblad operator (and so of the Hamiltonian for a dephasing channel). Therefore, these states do not contribute to the Fisher information (we can show that there exists a zero eigenvalue only for $p$ even).

First, let us assume that the eigenvalues are all different $0<\xi_{1}<\xi_n<\cdots<\xi_M$ where $\xi_1$ is the first positive eigenvalue (also denoted $\xi_m$), $\xi_M$ is the maximum eigenvalue. 
Keeping the leading order $\Lambda_1=\min_{\alpha}(\Lambda_\alpha)\neq 0$ in equation \eqref{evrho} we find another condition $\kappa_{1}^{(\alpha)}<\kappa_{2}^{(\alpha)}<\cdots<\kappa_{M-1}^{(\alpha)} < \kappa_{M}^{(\alpha)}$ as all the constant $\kappa_{n}^{(\alpha)}$ must be different from each other. In addition, we must have $\kappa_1^{(M)}>0$ as the sum over all $\kappa_1^{(n)}$ vanishes. The last properties imply that the two coefficients $c_M$ and $c_m$ in equation \eqref{SM:Coef} are of the order $e^{-\frac{\gamma}{2}\Lambda_1^2 t}$ as 
$$
c_M = \frac{(\kappa_{1}^{(M)}-\kappa_{1}^{(1)})e^{-\frac{\gamma}{2}\Lambda_1^2 t}+O\left(e^{-\frac{\gamma}{2}\Lambda_2^2 t}\right)}{1+O\left(e^{-\frac{\gamma}{2}\Lambda_1^2 t}\right)}\ ,\ c_m =  \frac{\min_{n,n'}\left\{(\kappa_{1}^{(n)}-\kappa_{1}^{(n')})\right\}e^{-\frac{\gamma}{2}\Lambda_1^2 t}+O\left(e^{-\frac{\gamma}{2}\Lambda_2^2 t}\right)}{1+O\left(e^{-\frac{\gamma}{2}\Lambda_1^2 t}\right)}
$$   

Now, if the eigenvalues are degenerated (which is the case for the Lindbladian \eqref{SM:pbodyDephasing}). For the upper bound, the results is the same than previously.  However, for the lower bound it turns out that $c_m=0$. To improve this bound, it suffices to notice that $\la \xi_M | H - \la H \ra_0 |\xi_M \ra = 0$ as the expectation value of the Hamiltonian is $\la H \ra_0 =\la \xi_M | H | \xi_M \ra$. Therefore following a similar argument than in section \ref{SM:QFIbound1} we find
\begin{align*}
\mathfrak{F}_x &\geq  \frac{4t^2}{\hbar^2}\sum_{n'} \xi_M \left(\frac{\xi_M-\xi_{n'}}{\xi_M+\xi_{n'}}\right)^2\left| \la \xi_M|H-\la H\ra_0 | \xi_{n'}\ra \right|^2, \\
&\geq c_m\frac{4t^2}{\hbar^2}\sum_{n'} \xi_M \left| \la \xi_M|H-\la H\ra_0 | \xi_{n'}\ra \right|^2, \\
&\geq c_m\frac{4t^2}{\hbar^2}\sum_{n'} \xi_M \left| \la \xi_M|H-\la H\ra_0 | \xi_{n'}\ra \right|^2, \\
&\geq c_m\frac{4t^2}{\hbar^2}\sum_{n'} \xi_M  \la \xi_M|H-\la H\ra_0 | \xi_{M}\ra, \\
& = c_m\frac{4t^2}{\hbar^2}\Delta H^2,
\end{align*}
where 
$$
c_m = \min_{\xi_{n'}}\left(\frac{\xi_M-\xi_{n'}}{\xi_M+\xi_{n'}}\right)^2 = \left(\frac{\xi_M-\xi_{M-1}}{\xi_M+\xi_{M-1}}\right)^2\ .
$$

Next, we compute the minimal non-zero value of 
$
\lambda_{ij}^2=\sum_{\nu}\left(\lambda_i^{(\nu)}-\lambda_j^{(\nu)}\right)^2\ .
$
To do this, we use the previous results in section \ref{SM:evSpin} giving the eigenvalues of the $p$-body Lindblad operator $L_{\nu}=\sigma^z_{i_1}\cdots\sigma^z_{i_p}$ with $\nu=\{i_1<\cdots<i_p\}$ as well as their degeneracy. 
We recall that the eigenvalues of the Lindblad operator $L_{\nu}$ are $\pm 1$, and so 
$$
\left(\lambda_i^{(\nu)}-\lambda_j^{(\nu)}\right)^2 = \left\{ \begin{array}{ll} 0 ,\ \text{if}\ i=j\ (\text{and}\ j=2^n+1-i\ \text{if k is even}) \\ 4 ,\ \text{otherwise}\end{array} \right.\ .
$$ 
Using the symmetry of the density matrix, we can pick $i=1$ associated with the vector $|e_0^{(0)}\ra = |--\cdots-\ra$ and compute $\lambda_{1j}$ for different values of $j=q+1$ varying from $2$ to $N+1$ associated with the vectors $|e_q^{(0)}\ra,\ q=1,2,\cdots,N$ defined in section \ref{SM:SpinONB}. The other values of $j$ associated with the other vectors $|e_q^{(m)}\ra$ are obtained by permutations of the values obtained for $j=q+1$. Using the degeneracy provided in section \ref{SM:evSpin} we find 
$$
\lambda_{1,q+1}^2 = \sum_{\nu:\lambda_{q+1}^{(\nu)}=-\lambda_1^{(\nu)}}4 =\sum_{s=0}^{(p-1)/2}{q\choose 2s+1}{N-q\choose p-(2s+1)}\ .
$$   
The minimal value (assuming $p<\lfloor \frac{N}{2} \rfloor$) obtained for $q=1$ (associated with the vector $|--\cdots-+\ra$) reads
$$ 
\Lambda_1^2 \equiv \min_{q=1,\cdots,N} \lambda_{1,q+1}^2 = 4{N-1\choose p-1} \sim \frac{4}{(p-1)!}N^{p-1}\ .
$$ 
This concludes the proof of the claim (see \textit{Product state and metrology equivalence} in the main body of the article)
$$
c_M\ \text{and}\ c_m\sim \exp{\left[-\frac{2\gamma}{(p-1)!}N^{p-1}t\right]}\ , 
$$
implying that the interrogation time scales as $t_{\text{int}}\sim \gamma^{-1}N^{-(p-1)}$. For $p=1$ we find that $t_{\text{int}}\sim\gamma^{-1}$ is independent of $N$ while it depends on $N$ for $p\geq 2$. 
Notice that the interrogation time has the scaling of the decoherence time $\tau_D$ multiplied by $N$. Indeed, it was shown in \cite{CBCdC16} that 
$$ 
\tau_D = \left(\gamma\sum_{\nu} \Delta_{\rho_0}L_\nu^2\right)^{-1} \ ,
$$
where for all $\nu$ we have
$$
\Delta_{\rho_0}L_\nu^2 = \text{tr}\left[L_\nu^2\rho_0\right] - \text{tr}\left[L_\nu\rho_0\right]^2 
= \la (\sigma_i^z)^2 \ra^p-\la \sigma_i^z \ra^{2p} = 1\ ,
$$
showing that 
$$ 
\tau_D = \gamma^{-1}{N\choose p}^{-1} \sim \gamma^{-1}p!\ N^{-p}\ .
$$
 
So far we considered only the pure-dephasing Lindbladian form \eqref{SM:pbodyDephasing}. Now, if one adds a $k$-body Hamiltonian that commutes with the $p$-body Lindblad operator $[H,L_\nu]=0$, we can prove that the solution of the master equation 
\begin{equation}\label{SM:MasterEq}
\dot{\rho}(t) = -\frac{ix}{\hbar}\left[H,\rho(t)\right] + \mathcal{D}_p(\rho)\ ,
\end{equation}
(here $x$ denotes the estimated parameter) is obtained from the unitary evolution of the previous density matrix (solution of the master equation with $H=0$) with $U(t,0) \equiv \exp{\left(-\frac{ix}{\hbar}H t\right)} $ and $U^\dagger(t,0)$
$$
\rho(t)\mapsto U(t,0)\rho(t)U^{\dagger}(t,0)\ .
$$
This shows that the eigenvalues of the density matrix $\rho(t)$ are not modified by the Hamiltonian. On the contrary the eigenvectors are modified as follows $|\xi_n\ra \mapsto U(t,0)|\xi_n\ra $ but that does not affect our conclusions above.  

Notice that the eigenstates of the density matrix do not depend on the parameter $\gamma$ while the eigenvalues are independent of $x$. This implies that the joint estimations of $x$ and $\gamma$ does not change the above results as the off-diagonal elements of the QFI matrix vanish.  We mention that this is specific to the master equation \eqref{SM:MasterEq} as in general the eigenstates of the density matrix depend on $\gamma$.

\section{Long-range Ising model with dephasing}\label{SM:LRIsing}

In the main body of the article, we want to estimate $x_1$ for the Long-range Ising model with $p$-body dephasing
\begin{equation}\label{SM:IsingDephasingNS}
\mathcal{L}(\rho)=-\frac{i}{\hbar}[H_I,\rho]+\mathcal{D}_p(\rho)\ ,
\end{equation}
where $H_I = -\sum_{i<j}J_{ij}\sigma_i^z\sigma_j^z$, with $J_{ij}=J/|i-j|^\alpha$, where the exponent $\alpha\geq 0$ controls the range of the interactions, and where the $p$-body dissipator is 
\begin{equation}\label{SM:pbodyDissipator}
\mathcal{D}_p(\rho)\equiv \gamma\sum_{i_1<\cdots<i_p}\left(\sigma_{i_1}^z\cdots\sigma_{i_p}^z\rho\sigma_{i_1}^z\cdots\sigma_{i_p}^z-\rho\right)\ .
\end{equation} 
In the main body of the paper we discuss the estimation of the coupling parameter $J$. This is more difficult than for the infinite range case $\alpha=0$ as for $\alpha>0$ the interactions depend on the distance between the sites. However, our formalism allows to treat this case as equation () shows that the QCR bound only depends on the initial variance of the Hamiltonian 
\begin{equation}\label{SM:H}
H\equiv -\sum_{i<j}\frac{\sigma_i^z\sigma_j^z}{|i-j|^\alpha} = \frac{1}{J}H_I\ .
\end{equation} 
In this section we want to give details of the derivation of the scaling of the variance of the Hamiltonian $H$ for both maximizing and product states.

\subsection{State that maximizes the variance of the Hamiltonian $H_I$}

For this model, the GHZ-state does not maximize the variance of the Hamiltonian $H$ as the ground state energy 
$$
E_m=-\sum_{i<j}\frac{1}{(j-i)^{\alpha}}\ ,
$$ 
has degenerate states $|-\cdots -\ra$ and $|+\cdots +\ra$. However, we can construct states that maximize the variance of the Hamiltonian, such as
$$ 
|\Psi_0\ra =  |E_m\ra + | E_M\ra = |\underbrace{-\cdots -}_{N}\ra + |\underbrace{-\cdots -}_{N-\lfloor \frac{N}{2}\rfloor }\underbrace{+\cdots +}_{\lfloor \frac{N}{2}\rfloor} \ra \ , 
$$
where $\lfloor x \rfloor $ is the floor function.
Here, the maximum energy is $E_M = \sum_{i<j}\varepsilon_{i,j}(j-i)^{-\alpha}$ where $\varepsilon_{i,j}=-1$ if $i,j\leq\ 
\text{or}\ \geq \lceil \frac{N}{2}\rceil$ and $+1$ otherwise. The seminorm of the operator $H$ is then $\|H\| = E_M-E_m = 2\sum_{i<\lfloor \frac{N}{2} \leq j}(j-i)^{\alpha}$.

$$
\Delta H \approx  \left\{ \begin{array}{ll}c_{\alpha}N^{2-\alpha} \ \ \ , \ 0\leq\alpha\leq 1 \\ \log{(N)} ,\ \alpha=2,\ \\ c_{\alpha},\ \alpha>2 \end{array} \right.\ ,
$$ 
where $c_{\alpha}=(1-2^{1-\alpha})((\alpha-2)(\alpha-1))^{-1}$ for $\alpha<1$, $\log{(2)},\ \alpha=1$, and $1+((\alpha-2)(\alpha-1))^{-1}$ for $\alpha>2$.

The eigenvalues $\lambda_{m}$ and $\lambda_M$ of the Lindblad operator $\sigma_i^z$ corresponding to the state $|E_m\ra$ and $|E_M\ra$ are respectively $-1$ and $-1$ for $i\leq \lfloor \frac{N}{2}\rfloor$ or $+1$ for $i\geq \lfloor \frac{N}{2}\rfloor$. Therefore, the sum of the variance square of the Lindblad oeprators $\sum_i (\Delta\sigma_i)^2$ scales as $\frac{N}{2}$. \\

\fbox{\begin{minipage}{50em}
To show that the state $| E_M\ra$ maximizes the eigenvalues of the Hamiltonian, we compute the eigenvalue associated with the states $|e_{q}^{(0)}\ra$ minus the ground state 
$$
\delta_q = \la e_{q}^{(0)}|H|e_{q}^{(0)}\ra - E_m = -\left(\sum_{i<j\leq N-q}\frac{1}{(j-i)^\alpha}+\sum_{N-q<i<j}\frac{1}{(j-i)^\alpha}\right)+\sum_{i\leq N-q<j}\frac{1}{(j-i)^\alpha} - E_m = 2\sum_{i\leq N-q<j}\frac{1}{(j-i)^\alpha}\ ,
$$
and maximize $\delta_q$ to obtain the seminorm $\|H\| = E_M-E_m$.
We estimate the last sum using Euler-Maclaurin formula
$$ 
\sum_{i\leq N-q<j}\frac{1}{(j-i)^\alpha} \approx 1+\int_{N-q+1}^{N}dy\int_{1}^{N-q} dx\ \frac{1}{(y-x)^\alpha} = 
\left\{ \begin{array}{ll} 1+\frac{1}{(2-\alpha)(1-\alpha)}\left[(N-1)^{2-\alpha}-(N-q)^{2-\alpha}-q^{2-\alpha}+1\right],\ \alpha\neq 1,2 \\ \\ N\ln{\left(\frac{N-1}{N-q}\right)}+q\ln{\left(\frac{N-q}{q}\right)}-\ln{\left(N-1\right)},\ \alpha=1 \\ \\ 1+\ln{\left(\frac{(N-q)q}{N-1}\right)},\ \alpha=2 \end{array} \right.\ ,
$$
which is maximized for $q = \lfloor \frac{N}{2} \rfloor$. It follows that the scaling of the seminorm given by
 $$ 
\|H\| \approx 
\left\{ \begin{array}{ll} \frac{1-2^{\alpha-1}}{(2-\alpha)(1-\alpha)}N^{2-\alpha},\ \alpha<1 \\ \\ N\ln{(2)},\ \alpha=1 \\ \\ \ln{(N)},\ \alpha=2 \\ \\ 1+\frac{1}{(2-\alpha)(1-\alpha)} \end{array} \right.\ .
$$ 

\end{minipage}}
\linebreak\\ \linebreak\\ 

\subsection{Product state and metrology equivalence} 
We now assume the initial state to be prepared in a product state $\rho_0 = \bigotimes_{i=1}^{N}\rho_0^{(1)}$ where the initial density matrix of an individual two-level system is chosen to be $\rho_0^{(1)} = |\psi\ra\la\psi|$ with $|\psi\ra=\left(\cos{(\phi)}|\varepsilon_m\ra +\sin{(\phi)}|\varepsilon_M\ra\right)/\sqrt{2}$ where the angle $0<\phi < \pi/2$.
We want to compute the variance of the Hamiltonian $H$ defined by equation \eqref{SM:H}. First let us write explicitly  the square of the Hamiltonian
$$
H^2 = \sum_{i<j}\frac{1}{|i-j|^{2\alpha}} + 2\sum_{i<j<k} \left\{\frac{\sigma_j^z\sigma_k^z}{|i-j|^\alpha |i-k|^{\alpha}}+\frac{\sigma_i^z\sigma_j^z}{|i-k|^\alpha |j-k|^{\alpha}}+\frac{\sigma_i^z\sigma_k^z}{|i-j|^\alpha |j-k|^{\alpha}}\right\} +\sum_{i<j\neq k<l\neq i}\frac{\sigma_i^z\sigma_j^z\sigma_k^z\sigma_l^z}{|i-j|^\alpha |k-l|^{\alpha}}\ ,
$$
where we assume that $N$ is large enough to find the last term. The mean value of $\sigma_i^z$ is
$$
\langle \sigma_i^z \rangle = \Big(\langle +|\cos(\phi) + \langle -|\sin(\phi) \Big) \Big(\cos(\phi)|+\rangle - \sin(\phi) |-\rangle \Big) = \cos^2(\phi) - \sin^2(\phi) = \cos(2\phi)\ ,
$$ 
hence we obtain
$$
\langle H^2 \rangle = \sum_{i<j}\frac{1}{|i-j|^{2\alpha}} + 2\cos^2(2\phi)\sum_{i<j<k}\left\{ \frac{1}{|i-j|^\alpha |i-k|^{\alpha}}+ \frac{1}{|i-k|^\alpha |j-k|^{\alpha}}+ \frac{1}{|i-j|^\alpha |j-k|^{\alpha}}\right\} +\cos^4(2\phi)\sum_{i<j\neq k<l\neq i} \frac{1}{|i-j|^\alpha |k-l|^{\alpha}}\ ,
$$
and for the expectation of the Hamiltonian we obtain a similar expression
\begin{multline*}
\langle H \rangle^2 = \left(\cos^2(2\phi) \sum_{i<j}\frac{1}{|i-j|^{\alpha}}\right)^2 \\
= \cos^4(2\phi) \left( \sum_{i<j}\frac{1}{|i-j|^{2\alpha}}+2\sum_{i<j<k}\left\{ \frac{1}{|i-j|^\alpha |i-k|^{\alpha}}+ \frac{1}{|i-k|^\alpha |j-k|^{\alpha}}+ \frac{1}{|i-j|^\alpha |j-k|^{\alpha}}\right\}+\sum_{i<j\neq k<l\neq i}\frac{1}{|i-j|^{\alpha}|k-l|^{\alpha}}\right)\ .
\end{multline*}
gathering the above results we find
\begin{equation}
\Delta H^2 \equiv \langle H^2\rangle - \langle H\rangle^2 = \left(1-\cos^4(2\phi)\right) \sum_{i<j}\frac{1}{|i-j|^{2\alpha}} + 2\left(\cos^2(2\phi)-\cos^4(2\phi)\right)\sum_{i<j<k}\left\{ \frac{1}{|i-j|^\alpha |i-k|^{\alpha}}+ \frac{1}{|i-k|^\alpha |j-k|^{\alpha}}+ \frac{1}{|i-j|^\alpha |j-k|^{\alpha}}\right\}\ ,
\end{equation}
where the second term is the leading term for large $N$ (as long as $\alpha$ is not too large). It reads
$$
\sin^2(4\phi)\sum_{i<j<k}\left\{ \frac{1}{|i-j|^\alpha |i-k|^{\alpha}}+ \frac{1}{|i-k|^\alpha |j-k|^{\alpha}}+ \frac{1}{|i-j|^\alpha |j-k|^{\alpha}}\right\}\ ,
$$
showing that it is maximal for $\phi = \pi/8$. Each sum above can be estimated using the Euler-Maclaurin  formula 
\begin{align*}
\sum_{i<j<k} \frac{1}{|i-j|^\alpha |i-k|^{\alpha}} &\approx \int_{3}^{N} dz\int_{2}^{z-1}dy\int_{1}^{y-1}dx\ \frac{1}{(y-x)^\alpha (z-x)^{\alpha}}\\
& \approx N^{3-2\alpha}\int_{0}^{1} dz'\int_{0}^{z'-1}dy'\int_{0}^{y'-1}dx'\ \frac{1}{(y'-x')^\alpha (z'-x')^{\alpha}}\ ,
\end{align*}
that shows the scaling of the sum  wth $N$ for $\alpha<\frac{3}{2}$.

\end{widetext}


\begin{thebibliography}{99}




%%%%%%%%%%%%%%%%%%%%%%%%%%%%%% Advanced in Quantum Metrology %%%%%%%%%%%%%%%%%%%%%%%%%%%%%%%%%%%%%%%%%%%

\bibitem{Giovannetti04} V. Giovannetti, S. Lloyd, L. Maccone,
% Quantum-Enhanced Measurements: Beating the Standard Quantum Limit
\href{http://science.sciencemag.org/content/306/5700/1330}{Science {\bf 306}, 1330 (2004).} 


\bibitem{Giovannetti11} V. Giovannetti, S. Lloyd, L. Maccone,
% Advances in quantum metrology
\href{http://www.nature.com/nphoton/journal/v5/n4/full/nphoton.2011.35.html}{Nature Photo. \textbf{5}, 222 (2011).} 

%%%%%%%%%%%%%%%%%%%%%%%%%%%%%% Gravitational wave %%%%%%%%%%%%%%%%%%%%%%%%%%%%%%%%%%%%%%%%%%%

\bibitem{Caves81} C. M. Caves, 
% Quantum-mechanical noise in an interferometer,
\href{https://doi.org/10.1103/PhysRevD.23.1693}{Phys. Rev. D {\bf 23,} 1693 (1981). }

\bibitem{LIGO11} The LIGO Scientific Collaboration, 
% A gravitational wave observatory operating beyond the quantum shot-noise limit,
\href{http://www.nature.com/nphys/journal/v7/n12/full/nphys2083.html}{Nature Phys. {\bf 7}, 962 (2011).}


%%%%%%%%%%%%%%%%%%%%%%%%%%%%%% Atomic spectroscopy and clocks %%%%%%%%%%%%%%%%%%%%%%%%%%%%%%%%%%%%%%%%%%%

\bibitem{Wineland92} D. J. Wineland, J. J. Bollinger, W. M. Itano, F. L. Moore, and D. J. Heinzen,
% Spin squeezing and reduced quantum noise in spectroscopy
\href{https://doi.org/10.1103/PhysRevA.46.R6797}{Phys. Rev. A {\bf 46}, R6797 (1992).} 

\bibitem{Bollinger96} J. J. Bollinger, W. M. Itano, D. J. Wineland, and D. J. Heinzen, 
% Optimal frequency measurements with maximally correlated states
\href{https://doi.org/10.1103/PhysRevA.54.R4649}{Phys. Rev. A {\bf 54}, R4649 (1996).} 

\bibitem{Giovannetti01} V. Giovannetti, S. Lloyd, L. Maccone,
% Quantum-enhanced positioning and clock synchronization
\href{http://www.nature.com/nature/journal/v412/n6845/full/412417a0.html}{Nature {\bf 412}, 417 (2001).} 


%%%%%%%%%%%%%%%%%%%%%%%%%%%%%%%%%%%%%%%%%%%%%%%%%%%%%%%%%%%%%%%%%%%%%%%%%%%%%%%%%%%%%%%%%%
%%%%%%%%%%%%%%%%%%%%%%%%%%%%%%  Mathematical framework QMetro %%%%%%%%%%%%%%%%%%%%%%%%%%%%%%%%%%%%%%



\bibitem{Helstrom76} C. W. Helstrom, \textit{Quantum Detection and Estimation Theory} (Academic Press, New York, 1976).

\bibitem{Holevo82} A. S. Holevo, \textit{probabilistic and Statistical Aspects of Quantum Theory} (North-Holland, Amsterdam, 1982). 

\bibitem{Wiseman10} H. M. Wiseman and G. J. Milburn, \textit{Quantum Measurement and Control} (Cambridge University Press, 2010).   

\bibitem{Braunstein94} S. Braunstein and C. M. Caves, 
% Statistical Distance and the Geometry of Quantum States
\href{https://doi.org/10.1103/PhysRevLett.72.3439}{Phys. Rev. Lett. {\bf 72}, 3439 (1994).}

\bibitem{Braunstein96} S. Braunstein, C. M. Caves, and G. J. Milburn,
% Generalized Uncertainty Relations: Theory, Examples, and Lorentz Invariance
\href{http://www.sciencedirect.com/science/article/pii/S0003491696900408}{Ann. of  Phys. {\bf 247}, 135 (1996).}


%%%%%%%%%%%%%%%%%%%%%%%%%%%%%%%%%%%%%%%%%%%%%%%%%%%%%%%%%%%%%%%%%%%%%%%%%%%%%%%%%%%%%%%%%%
%%%%%%%%%%%%%%%%%%%%%%%%%%%%%% Noisy Quantum Metrology %%%%%%%%%%%%%%%%%%%%%%%%%%%%%%%%%%%

%%%%%%%%%%%%%%%%%%%%%%%%%%%%%%  Huelga et al (1) %%%%%%%%%%%%%%%%%%%%%%%%%%%%%%%%%%%%%%%%%%%

\bibitem{Huelga97} S. F. Huelga et al.,
% Improvement of Frequency Standards with Quantum Entanglement
\href{https://doi.org/10.1103/PhysRevLett.79.3865}{Phys. Rev. Lett. {\bf 79}, 3865 (1997).}

%%%%%%%%%%%%%%%%%%%%%%%%%%%%%%  Maccone et al %%%%%%%%%%%%%%%%%%%%%%%%%%%%%%%%%%%%%%%%%%%


\bibitem{Demkowicz14} R. Demkowicz-Dobrza\'{n}ski and L. Maccone, 
% Using Entanglement Against Noise in Quantum Metrology
\href{https://doi.org/10.1103/PhysRevLett.113.250801}{Phys. Rev. Lett. {\bf 113}, 250801 (2014).}

\bibitem{Demkowicz12} R. Demkowicz-Dobrza\'{n}ski, J. Kolody\'{n}ski and M. Guta, 
% The elusive Heisenberg limit in quantum enhanced metrology
\href{http://www.nature.com/articles/ncomms2067}{Nat. Commun. {\bf 3}, 1063 (2012).}

%%%%%%%%%%%%%%%%%%%%%%%%%%%%%%  Escher et al %%%%%%%%%%%%%%%%%%%%%%%%%%%%%%%%%%%%%%%%%%%

\bibitem{Escher11} B. M. Escher, R. L. de Matos Filho, and L. Davidovich, 
% General framework for estimating the ultimate precision limit in noisy quantum-enhanced metrology
\href{http://www.nature.com/nphys/journal/v7/n5/abs/nphys1958.html}{Nat. Phys. {\bf 7}, 406 (2011).}

\bibitem{Escher11bis} B. M. Escher, R. L. de Matos Filho, and L. Davidovich, 
% Quantum Metrology for Noisy Systems
\href{http://link.springer.com/article/10.1007/s13538-011-0037-y}{Braz. J. Phys. {\bf 41}, 229 (2011).}

\bibitem{Escher12} B. M. Escher, L. Davidovich, N. Zagury, and R. L. de Matos Filho,
% Quantum Metrological Limits via a Variational Approach 
\href{https://doi.org/10.1103/PhysRevLett.109.190404}{Phys. Rev. Lett. {\bf 109}, 190404 (2012).}

%%%%%%%%%%%%%%%%%%%%%%%%%%%%%%  Huelga et al (2) %%%%%%%%%%%%%%%%%%%%%%%%%%%%%%%%%%%%%%%%%%%


\bibitem{Chin12} A. W. Chin, S. F. Huelga, M. B. Plenio, 
% Quantum Metrology in Non-Markovian Environments
\href{https://doi.org/10.1103/PhysRevLett.109.233601}{Phys. Rev. Lett. {\bf 109}, 233601 (2012). }

\bibitem{Smirne16} A. Smirne et al.,
% Ultimate Precision Limits for Noisy Frequency Estimation
\href{https://doi.org/10.1103/PhysRevLett.116.120801}{Phys. Rev. Lett. {\bf 116}, 120801 (2016).}


%%%%%%%%%%%%%%%%%%%%%%%%%%%%%%% Quantum Metro with 1/N^{3/2} type of scaling %%%%%%%%%%%%%%%%%%%%%%

\bibitem{Chaves13} R. Chaves et al., 
% Noisy Metrology beyond the Standard Quantum Limit 
\href{https://doi.org/10.1103/PhysRevLett.111.120401}{Phys. Rev. Lett. {\bf 111}, 120401 (2013). }

\bibitem{Dur14} W. D\"{u}r, M. Skotiniotis, F. Fr\"{o}wis, and B. Kraus,
% Improved Quantum Metrology Using Quantum Error Correction
\href{https://doi.org/10.1103/PhysRevLett.112.080801}{Phys. Rev. Lett. {\bf 112}, 080801, (2014).} 

%%%%%%%%%%%%%%%%%%%%%%%%%%%%%%%%%%%%%%%%%%%%%%%%%%%%%%%%%%%%%%%%%%%%%%%%%%%%%%%%%%%%%%%%%%


%%%%%%%%%%%%%%%%%%%%%%%%%%%%%%%%%%%%%%%%%%%%%%%%%%%%%%%%%%%%%%%%%%%%%%%%%%%%%%%%%%%%%%%%%%
%%%%%%%%%%%%%%%%%%%%%%%%%%%%%%%% Non-linear Quantum Metrology %%%%%%%%%%%%%%%%%%%%%%%%%%%%%%%%%%%%

%%%%%%%%%%%%%%%%%%%%%%%%%%%%%%  Boixo et al k-body %%%%%%%%%%%%%%%%%%%%%%%%%%%%%%%%%%%%%%%%%%%

\bibitem{Boixo07} S. Boixo, S. T. Flammia, C. M. Caves,  J. M. Geremia,  
\href{http://dx.doi.org/10.1103/PhysRevLett.98.090401}{Phys. Rev. Lett. {\bf 98}, 090401 (2007).}

\bibitem{Boixo08} S. Boixo et al., 
% Quantum-limited metrology with product states
\href{https://doi.org/10.1103/PhysRevA.77.012317}{Phys. Rev. A {\bf 77}, 012317 (2008). }
 
 %%%%%%%%%%%%%%%%%%%%%%%%%%%%%%  Hall and Wiseman Quantum Info %%%%%%%%%%%%%%%%%%%%%%%%%%%%%%%%%%%%%%%%%%%
 
\bibitem{Hall12}  M. J. W. Hall and H. M. Wiseman, 
% Does Nonlinear Metrology Offer Improved Resolution? Answers from Quantum Information Theory
\href{https://doi.org/10.1103/PhysRevX.2.041006}{Phys. Rev. X {\bf 2}, 041006 (2012).}

%%%%%%%%%%%%%%%%%%%%%%%%%%%%%%  Rezakhani k-body %%%%%%%%%%%%%%%%%%%%%%%%%%%%%%%%%%%%%%%%%%%

\bibitem{Alipour14} S. Alipour, M. Mehboudi, and A. T. Rezakhani,  
\href{http://dx.doi.org/10.1103/PhysRevLett.112.120405}{Phys. Rev. Lett. {\bf 112}, 120405 (2014).}

\bibitem{Alipour15} S. Alipour, and A. T. Rezakhani,  
% Extended convexity of quantum Fisher information in quantum metrology
\href{https://doi.org/10.1103/PhysRevA.91.042104}{Phys. Rev. A {\bf 91}, 042104 (2015).}

\bibitem{Yousefjani16} R. Yousefjani, S. Salimi, and A. S. Khorashad,  
\href{http://dx.doi.org/10.1007/s11128-017-1596-9}{Quantum Inf. Process., {\bf 16}(6), 144 ( 2017).}

%%%%%%%%%%%%%%%%%%%%%%%%%%%%%%  Atoms and BEC %%%%%%%%%%%%%%%%%%%%%%%%%%%%%%%%%%%%%%%%%%%

\bibitem{Rey07} A. M. Rey, L. Jiang, and M. D. Lukin, 
% Quantum-limited measurements of atomic scattering properties
\href{https://doi.org/10.1103/PhysRevA.76.053617}{Phys. Rev. A {\bf 76}, 053617 (2007).}

\bibitem{Choi08} S. Choi, B. Sundaram,  
\href{https://doi.org/10.1103/PhysRevA.77.053613}{Phys. Rev. A {\bf 77,} 220501 (2008).}

\bibitem{Boixo08b} S. Boixo et al., 
% Quantum Metrology: Dynamics versus Entanglement
\href{https://doi.org/10.1103/PhysRevLett.101.040403}{Phys. Rev. Lett. {\bf 101}, 040403 (2008). }

\bibitem{Boixo09} S. Boixo et al., 
% Quantum-limited metrology and Bose-Einstein condensates
\href{https://doi.org/10.1103/PhysRevA.80.032103}{Phys. Rev. A {\bf 80}, 032103 (2009). }

%%%%%%%%%%%%%%%%%%%%%%%% Optical Lattice clock %%%%%%%%%%%%%%%%%%%%%%%%%%%%%%%%%%%%%

\bibitem{MartinThesis} M. J. Martin, Quantum Metrology and Many-Body Physics: Pushing the Frontier of the Optical Lattice Clock, PhD thesis, University of Colorado, Boulder, USA (2013). 

\bibitem{Martin13} M. J. Martin, et al., 
% A quantum many-body spin system in an optical lattice clock
\href{https://doi.org/10.1126/science.1236929}{Science {\bf 341}, 632 - 636 (2013)}

\bibitem{Ludlow15} A. D. Ludlow, M. M. Boyd, J. Ye, E. Peik, and P. O. Schmidt,
% Optical Atomic clocks
\href{http://link.aps.org/doi/10.1103/RevModPhys.87.637}{Rev. Mod. Phys. {\bf 87}, 637 (2015).}

%%%%%%%%%%%%%%%%%%%%%%%%%%%%%%%%%%%%%%%%%%%%%%%%%%%%%%%%%%%%%%%%%%%%%%%%%%%%%%%%%%%%%%%%%%
%%%%%%%%%%%%%%%%%%%%%%%%%%%%%% Quantum Simulation of many-body decoherence  %%%%%%%%%%%%%%

\bibitem{CBCdC16} A. Chenu, M. Beau, J. Cao, and A. del Campo, 
\href{https://doi.org/10.1103/PhysRevLett.118.140403}{Phys. Rev. Lett. {\bf 118, }140403 (2017).}


%%%%%%%%%%%%%%%%%%%%%%%%%%%%%%%%%%%%%%%%%%%%%%%%%%%%%%%%%%%%%%%%%%%%%%%%%%%%%%%%%%%%%%%%%%%%%
%%%%%%%%%%%%%%%%%%%%%%%%%%%%%%% Long-range Ising model experiments   %%%%%%%%%%%%%%%%%%%%%%%%%%%%%%%


\bibitem{Richerme14} P. Richerme et al., 
% Non-local propagation of correlations in quantum systems with long-range interactions
\href{http://www.nature.com/nature/journal/v511/n7508/full/nature13450.html}{Nature {\bf 511}, 198 (2014).}

\bibitem{Jurcevic14} P. Jurcevic et al., 
% Quasiparticle engineering and entanglement propagation in a quantum many-body system
\href{http://www.nature.com/nature/journal/v511/n7508/full/nature13461.html}{Nature {\bf 511}, 202 (2014).}

\bibitem{Bloch08} I. Bloch, J. Dalibard, and W. Zwerger,
 \href{https://doi.org/10.1103/RevModPhys.80.885}{Rev. Mod. Phys. {\bf 80}, 885 (2008).}

%%%%%%%%%%%%%%%%%%%%%%%%%%%%%%%%%%%%%%%%%%%%%%%%%%%%%%%%%%%%%%%%%%%%%%%%%%%%%%%%%%%%%%%%%%
%%%%%%%%%%%%%%%%%%%%%%%%%%%%%%  Open Quantum Systems: theory %%%%%%%%%%%%%%%%%%%%%%%%%%%%%%%%%%%%%%%%%%%


\bibitem{Lindblad76}  G. Lindblad,  \href{http://dx.doi.org/10.1007/BF01608499}{Comm. Math. Phys. {\bf 48}, 119 (1976).}

\bibitem{Breuer02} H. P. Breuer and F. Petruccione, {\it The Theory of Open Quantum Systems} (Oxford University Press, New York, 2002).

\bibitem{Lidar06} D.A. Lidar, A. Shabani, R. Alicki, \href{http://dx.doi.org/10.1016/j.chemphys.2005.06.038}{Chem. Phys.  {\bf 322},  82 (2006).}

\bibitem{SM} See Appendix for technical details. 


%%%%%%%%%%%%%%%%%%%%%%%%%%%%%%%%%%%%%%%%%%%%%%%%%%%%%%%%%%%%%%%%%%%%%%%%%%%%%%%%%%%%%%%%%% Joint estimation strategy %%%%%%%%%%%%%%%%%%%%%%

\bibitem{Humphreys13}  P. C. Humphreys, M. Barbieri, A. Datta, and I. A. Walmsley
% Quantum Enhanced Multiple Phase Estimation 
\href{https://doi.org/10.1103/PhysRevLett.111.070403}{Phys. Rev. Lett. {\bf 111} , 070403 (2013).}

\bibitem{Cheng14} J. Cheng, \href{https://doi.org/10.1103/PhysRevA.90.063838}{Phys. Rev. A {\bf 90,} 063838 (2014).}
% Quantum metrology for simultaneously estimating the linear and nonlinear phase shifts


%%%%%%%%%%%%%%%%%%%%%%%%%%%%%%  QSL %%%%%%%%%%%%%%%%%%%%%%%%%%%%%%%%%%%%%%%%%%%

\bibitem{delCampo13}
A. del Campo, I. L. Egusquiza, M. B. Plenio,  S. F. Huelga, 
\href{http://dx.doi.org/10.1103/PhysRevLett.110.050403}{\prl{110}, 050403 (2013).}



%%%%%%%%%%%%%%%%%%%%%%%%%%%%%%%%%%%%%%%%%%%%%%%%%%%%%%%%%%%%%%%%%%%%%%%%%%%%%%%%%%%%%%%%%%%%%
%%%%%%%%%%%%%%%%%%%%%%%%%%%%%%% Many-body stochastic noise   %%%%%%%%%%%%%%%%%%%%%%%%%%%%%%%

\bibitem{Milburn91} G. J. Milburn, \href{https://doi.org/10.1103/PhysRevA.44.5401}{Phys. Rev. A {\bf 44}, 5401 (1991).}

\bibitem{Budini01}  A. A. Budini, \href{http://dx.doi.org/10.1103/PhysRevA.64.052110}{Phys. Rev. A {\bf 64}, 052110 (2001).}


%%%%%%%%%%%%%%%%%%%%%%%%%%%%%%%%%%%%%%%%%%%%%%%%%%%%%%%%%%%%%%%%%%%%%%%%%%%%%%%%%%%%%%%%%%
%%%%%%%%%%%%%%%%%%%%%%%% Quantum simulation of open quantum systems %%%%%%%%%%%%%%%%%%%%%%%%%%%%%%%%%%%%%

\bibitem{Barreiro11} J. T. Barreiro et al., 
% An open-system quantum simulator with trapped ions
\href{http://www.nature.com/nature/journal/v470/n7335/full/nature09801.html}{Nature {\bf 470}, 486 (2011).}

\end{thebibliography}
\end{document}